%% file: main.tex
\documentclass[reprint, amsmath,amssymb, pra]{revtex4-2}

\usepackage{preamble}
\usepackage{subcaption}

\begin{document}

\input{macros}

\message{The column width is: \the\columnwidth}

\preprint{APS/123-QED}

\title{Quantum control methods for robust entanglement of trapped ions}

\author{C. H. Valahu$^{1,2}$, I. Apostolatos$^{1}$, S. Weidt$^{1, 3}$, \\ W. K. Hensinger$^{1, 3}$}

\affiliation{$^{1}$Sussex Centre for Quantum Technologies, University of Sussex, Brighton, BN1 9QH, UK}
\affiliation{$^{2}$QOLS, Blackett Laboratory, Imperial College London, London, SW7 2BW, UK}
\affiliation{$^{3}$Universal Quantum Ltd, Brighton, BN1 6SB, UK}

\date{\today}
             
\input{_sections/Abstract/abstract}
\maketitle
\input{_sections/Introduction/introduction}
\input{_sections/Section1/section1.tex}
\input{_sections/Section2/section2.tex}
\input{_sections/Section3/section3.tex}
\input{_sections/Conclusion/conclusion}
\input{_sections/Acknowledgments/acknowledgments}

\appendix
\input{_sections/AppendixB/appendixB.tex}
\input{_sections/AppendixC/appendixC.tex}

\input{_sections/AppendixD/appendixD.tex}
\input{_sections/AppendixE/appendixE.tex}
\input{_sections/AppendixF/appendixF.tex}

\bibliography{references}

\end{document}

%% file: macros.tex
\renewcommand{\dag}{^\dagger}
\newcommand{\bigO}{\mathcal{O}}
\newcommand{\fid}{\mathcal{F}}
\newcommand{\infid}{\mathcal{I}}
\newcommand*{\bvec}[1]{\vv{\bm{#1}}} 
\newcommand{\minus}{\scalebox{0.75}[1.0]{$-$}} 
\newcommand{\yb}{\ce{^{171}Yb^+}}
\newcommand{\ndot}{\dot{\bar{n}}}
\newcommand{\nbar}{\bar{n}}
\newcommand{\msgate}{M{\o}lmer-S{\o}rensen}

\newcommand*{\eg}{e.g.\@\xspace}
\newcommand*{\ie}{i.e.\@\xspace}
\newcommand*{\cf}{c.f.\@\xspace}
\newcommand*{\myref}{Ref.\@\xspace}

\newcommand{\citetemp}[1]{[#1]}

\newcommand\T{\rule{0pt}{.5ex}}       
\newcommand\B{\rule[-.5ex]{0pt}{0pt}} 

%% file: _sections/Abstract/abstract.tex
\begin{abstract}

A major obstacle in the way of practical quantum computing is achieving scalable and robust high-fidelity entangling gates. To this end, quantum control has become an essential tool, as it can make the entangling interaction resilient to sources of noise. Nevertheless, it may be difficult to identify an appropriate quantum control technique for a particular need given the breadth of work pertaining to robust entanglement. To this end, we attempt to consolidate the literature by providing a non-exhaustive summary and critical analysis. The quantum control methods are separated into two categories: schemes which extend the robustness to (i) spin or (ii) motional decoherence. We choose to focus on extensions of the $\sigma_x\otimes\sigma_x$ \msgate\ interaction using microwaves and a static magnetic field gradient. Nevertheless, some of the techniques discussed here can be relevant to other trapped ion architectures or physical qubit implementations. Finally, we experimentally realize a proof-of-concept interaction with simultaneous robustness to spin and motional decoherence by combining several quantum control methods presented in this manuscript.

\end{abstract}

%% file: _sections/Introduction/introduction.tex
\section{Introduction} \label{sec:introduction}

Trapped ions are a promising platform for quantum information processing and have achieved the highest recorded fidelities to date \cite{bermudez2017, gaebler2016, srinivas2021, Pino2021, akhtar2022}. Nonetheless, these results were achieved on smaller NISQ devices and scaling high fidelities to many qubits in large processors remains an important challenge. While quantum error correction alleviates this bottleneck, consistent errors below the fault-tolerant threshold ($10^{-2}$) are still required \cite{raussendorf2007}. More practical thresholds are placed near $10^{-3}$ and $10^{-4}$, since the number of physical qubits that encode a single logical qubit scales with the infidelity. Surpassing the fault-tolerant threshold is hindered by the qubit's inevitable coupling to its noisy environment. Therefore, fault-tolerance is partly a classical engineering challenge, since one can reduce the noise via hardware improvement (\eg low-noise electronics or better shielding). In some cases, however, upgrading the classical control hardware comes at a large manufacturing cost and higher experimental overhead. Fortunately, one can instead engineer quantum control methods to reduce the qubit's coupling to its noisy environment, usually at a smaller cost of additional fields and modulations.

Quantum control methods for robust entangling gates are prevalent in trapped ion platforms. Laser gates have already demonstrated impressive fidelities \cite{benhelm2008, gaebler2016, Pino2021, clark2021} and fast gate times \cite{schafer2018}. Hybrid laser-microwave schemes that make use of a single sideband transition and continuous dynamical decoupling achieve robustness to thermal noise, dephasing and noise in the control fields themselves \cite{bermudez2012, lemmer2013, tan2013}.  Laser-free implementations have gained traction due to scalability issues associated with laser beams. Near-field all-microwave approaches with oscillating magnetic field gradients \cite{ospelkaus2008} have also demonstrated high-fidelities and added robustness using dynamical decoupling \cite{harty2014, harty2016}. Recent works with oscillating gradients have reported record fidelities with laser-free $\sigma_z\otimes\sigma_z$ gates that are simultaneously robust to spin and motional decoherence \cite{sutherland2019, sutherland2020, srinivas2021}. 

Another experimental implementation of laser-free gates instead uses static magnetic field gradients, which offers promising advantages when scaling up quantum processors to many qubits \cite{lekitsch2017}. This implementation is unique in that a magnetically sensitive transition must be used in order to obtain strong spin-motion coupling \cite{mintert2001}. The interaction therefore naturally suffers from dephasing, since the encoded qubit is linearly coupled to environmental noise. Entangling gates must then rely on quantum control techniques to achieve high-fidelities, whereas other trapped-ion platforms may use them as an added feature. This immediately restricts the available classes of quantum control methods that can be used, as they must extend the coherence time by at least several orders of magnitude. 

There are several experimental demonstrations of all-microwave entangling gates with a static magnetic field gradient that use quantum control to improve the fidelity. For example, a $\sigma_z\otimes\sigma_z$ gate was demonstrated with pulsed dynamical decoupling to extend the spin's coherence time \cite{khromova2012, piltz2013, piltz2016a}. It was also shown that continuous dynamical decoupling can be applied to this interaction to improve the robustness to spin decoherence while preserving the resilience to motional errors \cite{valahu2021}. Alternatively, entanglement via a \msgate\ type interaction has been demonstrated by encoding the qubits in a decoherence-free subspace via continuous dynamical decoupling \cite{randall2015, weidt2016}. Along with these examples, there exists a substantial library of quantum control schemes which extend the robustness of entangling gates, each with their own tradeoffs and advantages. This naturally leads to the following question: is there a quantum control scheme that is better suited for a particular experimental system and set of requirements?

In this manuscript, we aim to consolidate the breadth of quantum control methods that can be found in the literature and offer a succinct summary and comparison. The following sections attempt to provide a unifying framework which allows us to compare schemes with common metrics. The work presented here should be used as both an overview and a guide towards selecting robust entangling schemes. The quantum control methods are broadly separated in two categories: section \ref{sec:robustness_spin_dephasing} discusses schemes that extend the robustness to spin decoherence, while section \ref{sec:robustness_motional_decoherence} presents methods to extend the motional robustness. Finally, in section \ref{sec:experimental_demonstration}, we demonstrate an experimental proof-of-concept of an interaction that is simultaneously robust to spin and motional decoherence, by combining several of the aforementioned quantum control protocols. Note that only schemes pertaining to laser-free QCCD architectures with a static magnetic field gradient are considered (such as the architecture proposed in Ref. \cite{lekitsch2017}). We further assume that the entangling gate is generated by a bichromatic $\sigma_x\otimes\sigma_x$ \msgate\ interaction \cite{sorensen1999, sorensen2000}. Nevertheless, the results of this manuscript can be extended to a wider range of architectures and we hope that the comparisons can be useful for other qubit hardware.

%% file: _sections/Section1/section1.tex
\section{Robustness to spin dephasing} \label{sec:robustness_spin_dephasing}

\input{_sections/Section1/introduction.tex}

\input{_sections/Section1/pulsed_dd.tex}

\input{_sections/Section1/continuous_dd.tex}

\input{_sections/Section1/ml_continuous_dd.tex}

\input{_sections/Section1/summary.tex}

%% file: _sections/Section1/introduction.tex
Spin dephasing arises from a qubit coupling to its noisy environment and is described by the following Hamiltonian,

\begin{equation} \label{eq:basic_sigmaz_noise}
H_{noise} = \beta_z(t) \sigma_z.
\end{equation}

The stochastic variable $\beta_z(t)$ describes random fluctuations of the qubit frequency. Provided that the noise is non-Markovian, \ie the Power Spectral Density (PSD) $S_z(\omega) = \int^{+\infty}_{-\infty} \langle \beta_z(0) \beta_z(\tau)\rangle e^{-i\omega \tau}d\tau$ exhibits non-zero temporal correlations, quantum control methods can be employed to dynamically decouple the qubit from noise and extend the spin's coherence time. To this end, a \msgate\ entangling interaction can be made robust to spin decoherence. In the following section, we discuss and compare three such quantum control methods: (i) Pulsed Dynamical Decoupling (PDD), in which $\pi$-pulses are interleaved throughout the gate evolution; (ii) Continuous Dynamical Decoupling (CDD), where a carrier transition is continuously driven; (iii) Multi-Level Continuous Dynamical Decoupling (MLCDD), where CDD is applied to a multi-level system. These schemes are compared with one another using the following metrics:

\begin{description}

\item[\textbf{Fidelity}] \mbox{}

The fidelity achievable by a quantum control method is arguably the most important metric. In what follows, we use the decay function of the spin's coherence as a proxy for the infidelity \cite{ball2016},

\begin{equation} \label{eq:infidelity_spin_dephasing}
\infid = \frac{1}{2}(1 - e^{-\chi(\tau)}),
\end{equation}
where $\tau$ is the gate duration. In this way, the effectiveness of a quantum control method in achieving high fidelity entangling gates is linked with its ability to extend the spin's coherence. The decay function $\chi(t)$ is determined from the system's Hamiltonian. The PSD $S_z(\omega)$ can be encorporated into equation \ref{eq:infidelity_spin_dephasing} by using a filter function formalism: the qubit acts like a filter whose transfer function in the frequency domain is computed from the quantum control sequence \cite{biercuk2009, biercuk2011, green2013, soare2014}. The decay function is then calculated from the overlap of the noise's PSD with the filter transfer function $F(\omega, t)$,

\begin{equation} \label{eq:chi_filter_function}
\chi(t) = \frac{1}{2\pi} \int^{+\infty}_{-\infty} S_z(\omega) \frac{F(\omega, t)}{\omega^2} d\omega.
\end{equation}

Since noise spectrums vary between experimental systems, it is more useful to compare the quality of a quantum control scheme's filter function. In what follows, we therefore identify the functional form of $F(\omega, t)$ for each scheme.

\item[\textbf{Gate duration}] \mbox{}

Increasing the efficacy of a quantum control method often involves increasing the power of a dynamical decoupling field or the number of pulses. In the case where the given total power budget is constrained, this would imply diverting power from the entangling gate fields, thus prolonging the gate duration. It is therefore important to characterize this trade-off between the efficacy of the dynamical decoupling method and the achievable gate duration.

\item[\textbf{Robustness to static shifts}] \mbox{}

The qubit is subject to slow parameter drifts which are modelled as a constant offset with the replacement $\beta_z(t) \rightarrow \beta_z$ in equation \ref{eq:basic_sigmaz_noise}. While the quantum control methods under discussion efficiently decouple the qubit from fast fluctuating noise (c.f. equation \ref{eq:chi_filter_function}), they also make the qubit robust to these static qubit frequency missets. In order to characterize the robustness, we aim to build an empirical model for each quantum control method, depicting the achievable fidelity for a given static shift. 

\item[\textbf{Calibration requirements}]  \mbox{}

While robustness can improve the fidelity of an entangling
gate in the presence of drifts, experimental sequences are eventually required to recalibrate the system’s parameters. The scheduling rate of calibrations is determined by the tolerable reduction in fidelity over time. The total calibration duration and the scheduling rate should be kept small to maximize the continuous runtime of the quantum processor \cite{riesebos2021}. Therefore, the duration of the calibration sequences have a direct impact on the processor's duty cycle. This introduces a trade-off between a quantum control method's complexity and the processor's available runtime. We quantify this trade-off by representing the calibrations of each scheme as a Directed Acyclic Graph (DAG) \cite{kelly2018}, from which the number of nodes and dependencies provide a proxy for the calibration complexity and duration.

\item[\textbf{Experimental overhead}]  \mbox{}

The final figure of merit is the experimental overhead of a particular quantum control scheme. This includes the required physical resources, the general complexity of the scheme both conceptually and physically, and the stringency of hardware requirements. Since it is difficult to find a common metric for the experimental overhead, this figure of merit should serve as a summary of future challenges that may arise when experimentally implementing a particular scheme.

\end{description}

%% file: _sections/Section1/pulsed_dd.tex
\subsection{Pulsed Dynamical Decoupling} \label{sec:pulsed_dd}

The first demonstrations of quantum control within the nuclear magnetic resonance community extended the spin's coherence time via the application of $\pi$ pulses \cite{carr1954, meiboom1958}. The seminal Hahn spin echo consists of a single $\pi$-pulse at half the duration $\tau/2$, which refocusses qubit frequency fluctuations oscillating at frequencies $\omega<2/(\tau)$ \cite{hahn1950}. It was later shown that interleaving many $\pi$-pulses during the evolution may extend the coherence time further \cite{viola1998, uhrig2007, piltz2013}. It was also found that performing pulses along alternating orthogonal bases could efficiently decouple the qubit from noise in all three axes, and similar techniques can mitigate the effects of imperfections in the decoupling pulses themselves \cite{maudsley1986,viola1999,khodjasteh2005, souza2012}. All together, this vast library of pulse sequences form what we refer to as \textit{Pulsed Dynamical Decoupling} (PDD).

PDD schemes may be used to extend the coherence of spins during an entangling interaction. For example, $\pi$-pulses interleaved during a $\sigma_z\otimes\sigma_z$ entangling gate have been shown to suppress both static and time-varying qubit frequency noise \cite{piltz2013, piltz2016a, srinivas2021}. Protecting a M{\o}lmer-S{\o}rensen evolution in a similar manner is, however, more difficult, as the dynamical decoupling pulses do not necessarily commute with the gate fields. The nature of the bichromatic interaction causes the spin and motion to remain entangled throughout the evolution, further complicating the timings of the $\pi$ pulses. In what follows, we outline a variety of gate schemes that combine PDD with a $\sigma_x\otimes\sigma_x$ MS interaction.

Earlier works assumed that adding $\pi$-pulses while the spin and motion are still entangled will inevitably damage the fidelity since they do not commute with the gate operation, which affects the subsequent phase space trajectory. This led to the development of DD protocols in which refocussing pulses are only applied when the spin and motion are disentangled \cite{jost2009, hayes2012, ballance2015, ballance2016}. In practice, a $k$-loop MS gate is implemented by choosing a detuning $\delta = 2\sqrt{k}\epsilon \Omega_0$, with $\epsilon$ the Lamb-Dicke parameter and $\Omega_0$ the gate fields' Rabi frequency, and $\pi$-pulses are added at the completion of each loop. 

The PDD protocols can be made compatible with classical digital circuitry by building the pulse timings as multiples of a clock cycle \cite{hayes2011, ball2015, qi2017}. The timings are encoded by Walsh functions, which are a series of square pulses with values $\{-1, +1\}$, where $\pi$-pulses are applied during zero-crossings. These are complementary to a $k$-loop MS gate since the duration of a single loop can be viewed as a single clock cycle. The maximum number of $\pi$ pulses is then $N_\pi = k-1$. The gate speed suffers by a reduction factor of $\sqrt{k}$, such that $\tau = \sqrt{k}\tau_0$. It is therefore difficult to implement a large number of pulses due to the poor gate time scaling. Note that the application of a $\pi$-pulse when spin and motion are disentangled still alters the resulting phase space trajectory, usually causing a symmetry of the path about the origin. This was found to also increase motional robustness \cite{hayes2012}, and is further discussed in section \ref{sec:robustness_motional_decoherence}. 

There exist PDD sequences with more complex timings that are difficult to encode with Walsh functions (\eg Uhrig \cite{uhrig2007}). Nevertheless, combining these sequences with the MS interaction is in principle possible. The duration of a single loop can be altered by varying the MS fields' detuning and Rabi frequency. One could therefore imagine a pulse sequence that implements a $k$-loop MS gate with non-constant durations which match the timings of the PDD pulses. The total entangling phase picked up by all loops should then be identical to that of a primitive \msgate\ gate. 

It was later shown that $\pi$-pulses can be applied at any point in the MS evolution, even if the spin and motion remain entangled \cite{manovitz2017}. A $\pi$-pulse effectively flips the sign of the operators in the \msgate\ unitary. In phase space, this is equivalent to reversing the direction of the trajectory. The timings of the pulses can therefore be engineered such that fast reversals of the phase space trajectory disentangle the spin and motion at the gate time. The resulting trajectory traces out what is commonly referred to as a flower, where a $\pi$-pulse is added at every intersection of two petals (see Ref. \cite{manovitz2017} for a detailed explanation). The duration between two pulses required to implement a periodic PDD sequence is $\pi(2 + N_\pi)/N_\pi\delta$, where $N_\pi$ are the number of $\pi$-pulses. For a given MS detuning $\delta$, the gate time is chosen such that the flower trajectory encloses an area equal to the maximal entangling phase $\pi/2$. In the limit of many $\pi$-pulses, the total required gate time is $\tau = \frac{\pi}{2}\tau_0$ \cite{manovitz2017}. The advantage of this scheme therefore comes from the gate time scaling which is independent of the number of pulses and allows for fast PDD gates (assuming instantaneous $\pi$-pulses; see section \ref{sec:pdd_gate_speed} for a discussion on non-instantaneous pulses). Note that the detuning of the gate fields, the $\pi$-pulse timings and the gate duration can be optimized to implement complex PDD sequences. This scheme also allows for dynamical decoupling sequences with multi-axis pulses, such as the XY-4 or XY-8 schemes \cite{wang2012, farfurnik2015}.

\subsubsection{Fidelity}

As outlined in equations \ref{eq:infidelity_spin_dephasing} and \ref{eq:chi_filter_function}, the infidelity under PDD is estimated from the spin's loss of coherence. The filter function of an arbitrary sequence is \cite{biercuk2009}

\begin{align} \label{eq:pdd_filter_function}
& F_{PDD}(\omega, \tau) = \nonumber \\
& |1 + (-1)^{N_\pi +1}e^{i\omega\tau} + 2\sum^{N_\pi}_{j=1}(-1)^{j}e^{i\delta_j\omega\tau}\cos(\omega\tau_\pi/2)|^2,
\end{align}

where $N_\pi$ is the number of $\pi$-pulses, $\delta_j$ are their normalized timings and $\tau_\pi$ is the duration of a single pulse. In general, the PDD filter function corresponds to a high-pass filter, whose low-frequency roll-off largely varies between specific sequences, \cite{biercuk2011}. An efficient PDD sequence is one that results in a steeper roll-off. Furthermore, the corner frequency of the high-pass filter tends to increase with the number of pulses. A large number $N_\pi$ is therefore desirable such that the stop-band region of the filter function suppresses most of the noise. The resulting infidelity from dephasing is, to first order,

\begin{equation} \label{eq:pdd_infidelity_dephasing}
\infid = \frac{1}{4\pi}\int^{+\infty}_{-\infty}S_z(\omega) \frac{F_{PDD}(\omega, \tau)}{\omega^2}d\omega.
\end{equation}

\begin{figure}[t!]
	\center
    \includegraphics[scale=1.]{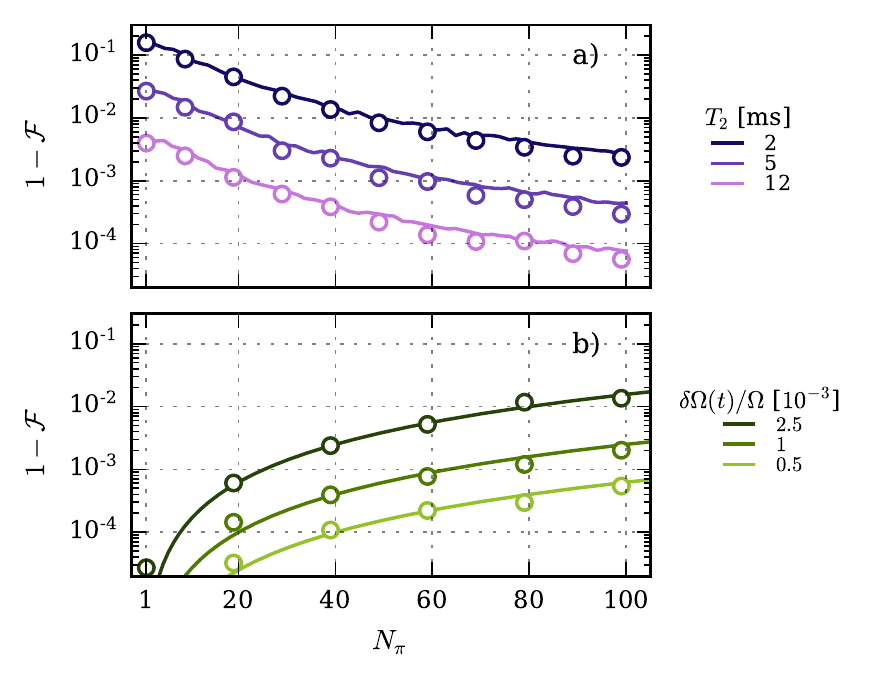}
    \caption{Numerical simulations of the \msgate\ entangling gate protected by pulsed dynamical decoupling and subject to spin dephasing noise. The circles represent the simulated infidelities averaged over 200 noise trajectories, in which qubit frequency fluctuations of equation \ref{eq:basic_sigmaz_noise} are modelled as an Orstein-Uhlenbeck process \cite{wang1945, gillespie1996, lemmer2013}. Solid lines are predictions from the analytical infidelity models of equations \ref{eq:pdd_infidelity_dephasing} and \ref{eq:pdd_infidelity_pulse_err}. (a) Infidelities due to qubit frequency fluctuations as a function of the number of noise-free pulses $N_\pi$. The coherence times $T_2$ are those of the magnetic sensitive transition, and are chosen such that $T_2 = \SI{2}{ms}$ is achievable on readily available systems, while $T_2 = \SI{12}{ms}$ can be obtained via additional noise shielding \cite{wang2021}. (b) Infidelities from pulse imperfections, where amplitude noise is injected into the refocussing pulses and the MS evolution is made noise-free. Relative amplitude noise levels were chosen from \cite{arrazola2020}.}
    \label{fig:pdd_noise_robustness}
\end{figure}

We identify another source of infidelity arising from imperfections in the dynamical decoupling pulses themselves. This error is evaluated by following a similar reasoning as in Ref. \cite{wang2012}. Because the pulse durations are generally fast with respect to the qubit frequency fluctuations, \ie $\tau_\pi \ll T_2$, the time-dependent noise $\beta_z(t)$ can be modelled as a constant offset during rotations. The same argument is made for amplitude fluctuations. In this way, all sources of  noise are modelled as constant shifts and the rotations become instantaneous unitaries with a static error in either the azimuthal Bloch sphere angle or the rotation (polar) angle. The non-ideal $\pi$-pulse operator is 

\begin{equation}
U_{k} = \exp\big( - i (\pi + \epsilon_k) (\bm{S}\cdot \bvec{n})) \big),
\end{equation}

where $\epsilon_k$ models over- and under-rotations, $\bm{S}$ are the Pauli matrices and $\bvec{n}= (n_x, n_y, n_z)$ is the rotation axis. Only considering the dynamical decoupling pulses, the total rotation of $N$ pulses becomes

\begin{equation}
U = \prod^N_{i} U_{k_i}.
\end{equation}

This final result can be used to evaluate the infidelity from pulse imperfections,

\begin{equation} \label{eq:pdd_infidelity_pulse_err}
\infid = \frac{Tr(U U_0)}{Tr(U)Tr(U_0)}
\end{equation}


where $U_0$ is the ideal error-free operator, obtained by setting $\epsilon _k = 0$, $n_{x/y} = n_z = 0$ and $n_{y,x} = 1$. 

An estimate of the total infidelity can be found from equations \ref{eq:pdd_infidelity_dephasing} and \ref{eq:pdd_infidelity_pulse_err}. The accuracy of these results is verified by numerically simulating the \msgate\ interaction under the influence of dephasing noise. The results, reported in figure \ref{fig:pdd_noise_robustness}, show good agreement between predicted and simulated infidelities. A trade-off also becomes apparent: large numbers of pulses $N_\pi$ are desired to more efficiently decouple the qubit from low frequency noise, however this also leads to a an increased accumulation of error due to pulse imperfections.

\subsubsection{Gate duration} \label{sec:pdd_gate_speed}

The shortest entangling duration that makes use of PDD is achieved by the fast gate scheme of Ref. \cite{manovitz2017}, which applies $\pi$-pulses while spin and motion remain entangled. The gate duration must be slightly prolonged given that the enclosed area in phase space is reduced. We recall that, in the limit of many $\pi$-pulses, the primitive gate duration $\tau_0$ increases by a factor of $\frac{\pi}{2}$. Along with the finite $\pi$-pulse durations, the total gate duration becomes 

\begin{equation}
\tau_{PDD} = \frac{\pi}{2}\tau_0 + N_\pi \tau_\pi.
\end{equation}

We note that the duration of the fast PDD scheme can, in principle be shortened by employing modulation techniques which alter the phase space trajectory (c.f. Section \ref{sec:sideband_modulation}). For example, instantaneous phase shifts in the bichromatic fields alters the MS motional phase, which subsequently changes the direction of the phase space trajectory. By appending such a modulation to every $\pi$-pulse, the phase space trajectory could be engineered to maximize the enclosed area. If the enclosed area corresponds to that of a primitive sequence, \ie a circle, the gate duration becomes 

\begin{equation}
\tau_{PDD} = \tau_0 + N_\pi \tau_\pi.
\end{equation}

This result places an additional trade-off on the number of pulses $N_\pi$. While many $\pi$-pulses are desired to improve the filtering properties (c.f. equation \ref{eq:pdd_filter_function}), this directly increases the gate duration, making the interaction more sensitive to other sources of noise, such as motional decoherence.

\subsubsection{Robustness to static shifts} \label{sec:pdd_static_robustness}

\begin{figure}[t!]
	\center
    \includegraphics[scale=1.]{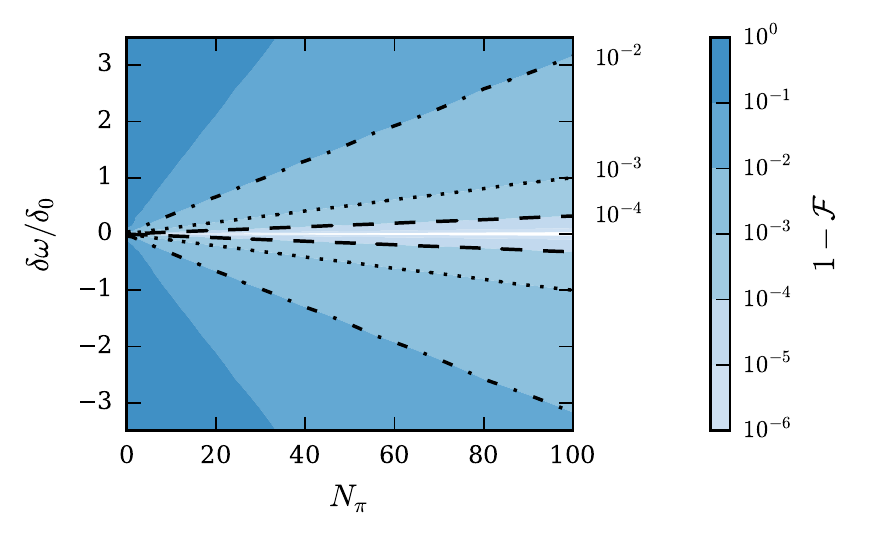}
    \caption{Robustness of the \msgate\ entangling gate protected by pulsed dynamical decoupling to static qubit frequency shifts. The Bell state fidelities are numerically simulated for a range of $\pi$-pulse numbers $N_\pi$ and normalised shifts $\delta\omega/\delta_0$. The dashed, dotted and dash-dotted lines are contours corresponding to the infidelities $10^{-4}$, $10^{-3}$ and $10^{-2}$.}
    \label{fig:pdd_static_robustness}
\end{figure}

Robustness of the PDD scheme to static qubit transition frequency shifts is investigated by means of numerical simulations. The noise Hamiltonian (equation \ref{eq:basic_sigmaz_noise}) generalized to multiple ions is integrated with the standard MS Hamiltonian after replacing the time-dependent noise by a static shift $\delta\omega$. A periodic PDD sequence is integrated by applying equally spaced $\pi$-pulses that are error-free and instantaneous. Figure \ref{fig:pdd_static_robustness} reports the Bell state fidelity for a range of pulse numbers $N_\pi$ and normalized detuning errors $\delta\omega/\delta_0$, where $\delta_0$ is the MS detuning. An empirical model of the robustness is constructed by fitting the contours of figure \ref{fig:pdd_static_robustness} to a linear function, and one finds

\begin{align} \label{eq:pdd_static_robustness}
\delta\omega/\delta_0 \  \leq 
\begin{cases}
 \ (2.8 +  3.2N_\pi)\times 10^{-2},  \  & \infid \leq 10^{-2},  \\
 \ (0.8 + N_\pi) \times 10^{-2},  &\infid \leq 10^{-3},  \\
 \ (0.3 + 0.3 N_\pi) \times 10^{-2}, & \infid \leq 10^{-4}. 
\end{cases}
\end{align}

The model of equation \ref{eq:pdd_static_robustness} estimates the tolerable qubit frequency shift that still allows infidelities below a certain threshold. For example, one can achieve infidelities below $10^{-3}$ despite a shift of $\delta\omega = 0.108\delta_0$ if $N_\pi=10$ pulses are chosen. Alternatively, a larger shift of $\delta\omega = 1.008\delta_0$ is possible if one chooses $N_\pi=100$ pulses. For a gate duration of $\tau_0 = \SI{1}{ms}$, this would correspond to $\delta\omega/2\pi \approx \SI{1}{kHz}$.

\subsubsection{Calibration requirements} \label{sec:pdd_calibration_requirements}

A calibration involves the determination of a parameter by fitting the results of an experiment. The results of a calibration experiment may be skewed by another parameter that is mis-calibrated. This introduces the notion of dependency, \ie some calibrations should precede others. A calibration sequence is a collection of calibration experiments that should be executed in a specific order to take into account any possible dependencies of a gate scheme. In this way, a calibration sequence can be modelled as a Directed Acyclic Graph (DAG), wherein vertices (or nodes) designate a parameter and arcs (edges) represent a dependency \cite{kelly2018}. We also make the distinction between weak and strong dependencies. In the case of a weak (strong) dependency, the child parameter's calibration is (in)valid despite a small misset in the parent vertex. 

\begin{figure}[t!]
    \centering
    \includegraphics[scale=1.]{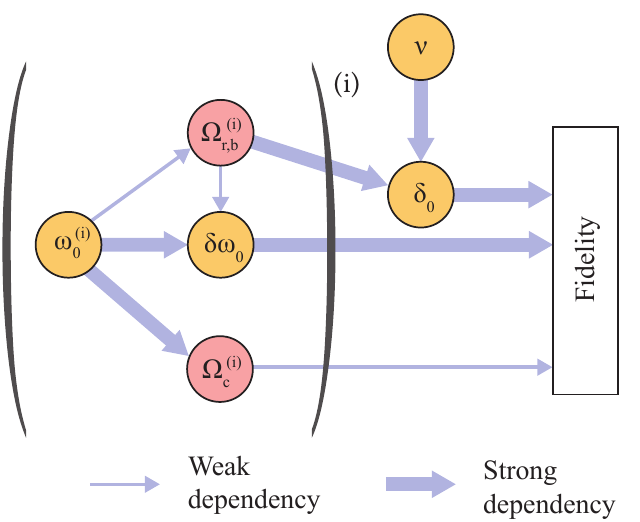}
    \caption{Directed acyclic graph describing the calibration requirements of both the pulsed and continuous dynamical decoupling schemes. The nodes represent parameters that need to be calibrated, and vertices represent dependencies. Yellow (red) nodes consist of transition frequencies (Rabi rates). Thin (thick) arrows represent weak (strong) dependencies. Nodes within parentheses indicate that this sub-graph should be repeated for every $i^{th}$ ion.}
    \label{fig:pdd_cdd_calibrations}
\end{figure}

The PDD scheme necessitates three fields per ion: a pair of sidebands for the bichromatic interaction and a single carrier for dynamical decoupling. Each field is parametrized by a transition frequency $\omega$, a Rabi frequency $\Omega$ and a phase $\phi$. Calibrations of the field's phases are not considered here, as typical microwave sources exhibit low phase drift and a high resolution. The resulting DAG for the PDD scheme is illustrated in figure \ref{fig:pdd_cdd_calibrations}. The graph contains calibrations for the qubit transition frequency $\omega_0$, the bichromatic and carrier Rabi frequencies $\Omega_{r,b}$ and $\Omega_c$, a possible stark shift $\delta\omega_0$, the MS detuning $\delta_0$ and finally the secular frequency $\nu$. In total, we identify 12 nodes, with 12 strong and 10 weak dependencies.


\subsubsection{Experimental overhead} \label{sec:pdd_experimental_req}

To make use of the magnetic gradient induced coupling scheme \cite{mintert2001}, the qubits should be encoded in a magnetic sensitive transition, therefore any of the $\ket{F=0, m_f=0}\rightarrow \ket{F=1, m_f=\pm 1}$ transitions are suitable. Both the MS and the dynamical decoupling interaction are performed on the hyperfine transition, and all frequencies are centered around $\SI{12.6}{GHz}$ in a bandwidth of 10 MHz. The PDD scheme is therefore an all microwave approach. Since all interactions are performed on the same hyperfine transition, only a single microwave polarisation of $\sigma^+$ or $\sigma^-$ is required. Note that during idle memory operations, it may be preferable to encode the qubit within the magnetic insensitive clock transition. If this is the case, an additional $\pi$-polarised field is required to map populations in and out of the memory qubit before and after a gate operation.

\begin{figure}[t!]
    \centering
    \includegraphics[scale=1.]{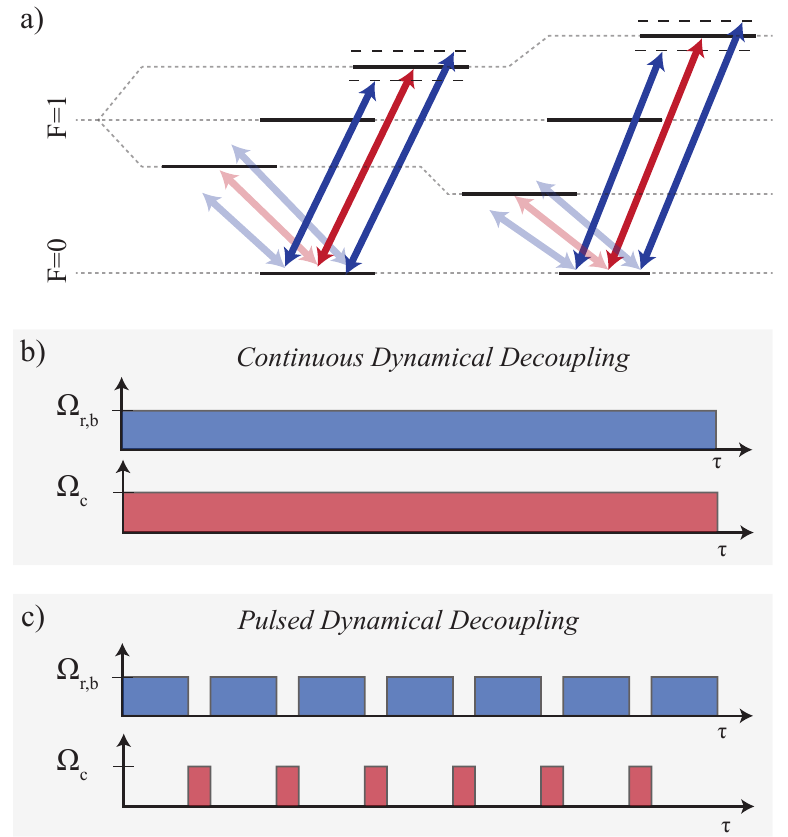}
    \caption{Quantum control scheme for PDD and CDD. Sideband (carrier) fields are in blue (red). (a) Energy levels of a typical hyperfine ground state. Ions have different resonance frequencies due to the static magnetic field gradient. PDD and CDD fields must address a magnetic sensitive transition between F=0 and F=1 (either of the opaque or transparent fields are possible). (b-c) Pulse sequences for both schemes. In CDD (b), both sidebands and carrier fields are continuously applied. In PDD (c), the carrier field is interleaved throughout the evolution.}
    \label{fig:pdd_cdd_schematic}
\end{figure}

Leakage to the nearest clock state transition, $\ket{F=0, m_f=0} \rightarrow \ket{F=1, m_f =0}$, may occur for larger microwave powers. Nevertheless, this transition is typically over $\SI{10}{MHz}$ away, leading to negligible couplings. Furthermore, cross-talk between ions in a magnetic field gradient is negligible \cite{piltz2014}. 

In principle, the gate could be performed within the $\ket{F=1}$ triplet states, i.e. on any of the $\ket{F=1, m_f = 0}\rightarrow \ket{F=1, m_f = \pm 1}$. This would allow for an all RF entangling gate. However, the frequency separations between both transitions due to the second order Zeeman shift are on the order of $\SI{10}{kHz}$ to $\SI{20}{kHz}$. Off-resonant coupling of the carrier and sideband transitions would therefore lead to appreciable infidelities, making the all RF approach impractical.

%% file: _sections/Section1/continuous_dd.tex
\subsection{Continuous Dynamical Decoupling} \label{sec:continuous_dd}

The previous section showed how a concatenation of $\pi$-pulses can effectively decouple the qubit from spin dephasing noise during a M{\o}lmer-S{\o}rensen evolution. An interesting case arises in the limit of large pulse times, \ie the dynamical decoupling drive is continuously applied during the entangling interaction. We refer to this as \textit{Continuous Dynamical Decoupling} (CDD). CDD has been shown to extend the coherence of spins in numerous platforms \cite{facchi2004, fanchini2007, cai2012, yan2013}.  Applying this dynamical decoupling method to trapped ion entangling gates was first proposed by utilizing a single red-sideband with a carrier \cite{bermudez2012, lemmer2013, tan2013}. In this way, the interaction relies on the carrier for both the dynamical decoupling and the entangling mechanism. This same idea was applied to the standard MS scheme, such that the interaction does not rely on the carrier drive, which now only provides additional dynamical decoupling \cite{harty2016}.

In order to elucidate the beneficial effects of the drive, we consider the usual MS Hamiltonian along with a field resonant with each qubit carrier transition and a noise term,

\begin{align}
& H  = H_{MS} + H_{drive} + H_{noise} \label{eq:hamiltonian_cdd}\\
& H_{drive} = \frac{\hbar \Omega_c}{2}\sum_i \sigma^{(i)}_x \\ 
& H_{noise} = \frac{\hbar\beta_z(t)}{2}\sum_i\sigma_z^{(i)}.
\end{align}

Provided that the basis of the drive is identical to that of the MS interaction, $H_{drive}$ and $H_{MS}$ commute with one another. Therefore, in an interaction basis with respect to the drive, \ie $\tilde{H} = e^{itH_{drive}/\hbar}(H-H_{drive})e^{-itH_{drive}/\hbar}$, $H_{MS}$ is unaffected and equation \ref{eq:hamiltonian_cdd} becomes

\begin{align}
& \tilde{H} = H_{MS} + \tilde{H}_{noise}, \\
& \tilde{H}_{noise} = \frac{\hbar\beta_z(t)}{2} \sum_i \left[ \cos(\Omega_c t)\sigma_z^{(i)} + \sin(\Omega_ct)\sigma_y^{(i)} \right].
\end{align}

The addition of a resonant drive therefore continuously rotates the qubit frequency noise $\beta_z(t)$ around the z and y axes. The terms of $\tilde{H}_{noise}$ are neglected under a rotating wave approximation in the limit $\beta_z(t) \ll \Omega_c$. The drives decouple the qubits from frequency fluctuations, and only noise at a frequency nearing $\Omega_c$ will decohere the spins. It is generally beneficial to introduce a large carrier Rabi frequency to better suppress qubit frequency fluctuations. We note that the interaction picture of the carrier drive must coincide with that of the MS evolution at the completion of the entangling gate. To this end, the carrier should perform an integer number of rotations during the evolution, which places a constraint on its Rabi freuqency, $\Omega_c = 2\pi k /\tau_0$ where $k \in \mathbb{Z}^+$.

\subsubsection{Fidelity}

\begin{figure}[t!]
	\center
    \includegraphics[scale=1.]{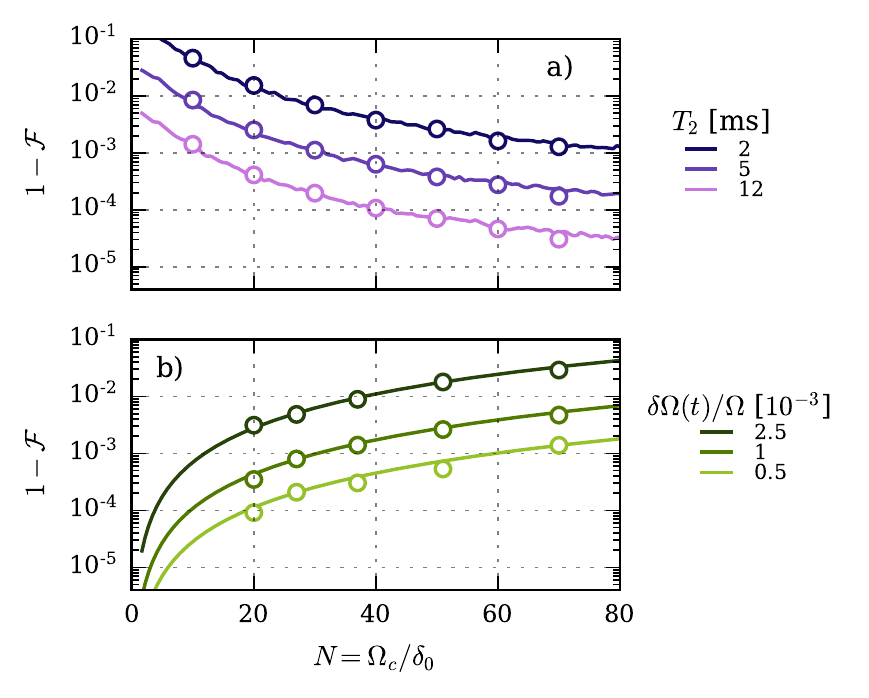}
    \caption{Numerical simulations of the \msgate\ entangling gate protected by continuous dynamical decoupling and subject to spin dephasing noise. (a-b) as captioned in figure \ref{fig:pdd_noise_robustness}, namely, numerical simulations used the exact Hamiltonian of equation \ref{eq:hamiltonian_cdd}, while analytical fidelities were obtained from equation \ref{eq:infidelity_cdd_dephasing} and \ref{eq:infidelity_cdd_amp_noise}. (a) Infidelities due to qubit frequency fluctuations for varying carrier Rabi frequencies, parametrized by the number of $2\pi$ pulse performed at the gate duration $N = \Omega_c/\delta_0$. The dynamical decoupling field is noise-free. (b) Infidelities from amplitude noise in the carrier dynamical decoupling fields, modelled by the replacement $\Omega_c \rightarrow \beta_x(t)\Omega_c$, where $\beta_x(t)$ is an Ornstein-Uhlenbeck process \cite{wang1945, gillespie1996, lemmer2013}. Here, dephasing noise is ignored by setting $\beta_z(t) =0$.}
    \label{fig:cdd_noise_robustness}
\end{figure}

The infidelity is estimated from the decay of the spin's coherence. From equations \ref{eq:infidelity_spin_dephasing} and \ref{eq:chi_filter_function}, we aim to find a filter function for the continuous drive. In an interaction picture with respect to $H_{noise}$, dephasing noise is suppressed by the energy gap that is opened in the new eigenstates. One then expects a sinc-like filter function, which consists of a narrow-bandwidth band-pass filter (see appendix \ref{app:filter_functions}). This is approximated by a Dirac delta function \cite{yan2013} and the infidelity becomes, to first order,

\begin{equation} \label{eq:infidelity_cdd_dephasing}
\infid = \frac{S_z(\Omega_c) \tau}{4}.
\end{equation}

An additional infidelity term arises when considering noise in the dynamical decoupling carrier field itself. Amplitude fluctuations introduce Rabi frequency noise, which is modelled as $\Omega_c\beta_x(t)\sigma_x$, where $\beta_x(t)$ is the fractional Rabi frequency noise with PSD $S_x(\omega) = \int^{+\infty}_{-\infty}\langle\beta_x(0)\beta_x(\tau)\rangle e^{-i\omega\tau}d\tau$. The noise amplitude therefore increases with $\Omega_c$, which limits the efficacy of the dynamical decoupling and introduces a trade-off. The infidelity from this additional noise is, from equation \ref{eq:infidelity_spin_dephasing} 

\begin{equation} \label{eq:infidelity_cdd_amp_noise}
\infid = \frac{\Omega_c^2}{2\pi} \int^{+\infty}_{-\infty} d\omega S_x(\omega) \frac{F_x(\omega,\tau)}{\omega^2}.
\end{equation}

With no further dynamical decoupling, the filter function $F_x(\omega,t)$ coincides with that of a free induction decay sequence, \ie $F_x(\omega, t) = 2\sin^2(\frac{\omega t}{2})$. A trade-off becomes apparent from equations \ref{eq:infidelity_cdd_dephasing} and \ref{eq:infidelity_cdd_amp_noise}. A large carrier Rabi frequency is desired to filter higher frequencies of noise which, for a typical correlated noise spectrum, contain weaker powers of noise. However, larger Rabi frequencies may increase the infidelity term of equation \ref{eq:infidelity_cdd_amp_noise}. This places a practical limit on the efficacy of continuous dynamical decoupling in the limit of larger Rabi frequencies. The infidelities of equations \ref{eq:infidelity_cdd_dephasing} and \ref{eq:infidelity_cdd_amp_noise} are verified by numerically integrating the full Hamiltonian of equation \ref{eq:hamiltonian_cdd}. The analytical model and simulation results show good agreement, and further illustrate the trade-off in the choice of the carrier Rabi frequency.

There exist various extensions to the CDD scheme that make the filter function of equation \ref{eq:infidelity_cdd_amp_noise} more efficient at decoupling the spins from amplitude noise. A first scheme consists in repeatedly adding dynamical decoupling drives with decreasing powers \cite{cai2012}. The addition of a second drive effectively opens a new energy gap which decouples the first drive’s amplitude noise. In principle, any number of concatenated drives can be implemented, each mitigating the noise introduced by the previous one. A different approach uses continuous phase modulation on the principle dynamical decoupling carrier field \cite{cohen2016, farfurnik2017}. Similarly to the concatenated drives, this also opens an additional energy gap which decouples the qubit from amplitude noise. Alternatively, one can apply refocussing pulses within the dressed eigenbasis to mitigate noise in the drive itself \cite{genov2019}.

\begin{figure}[t!]
	\center
    \includegraphics[scale=1.]{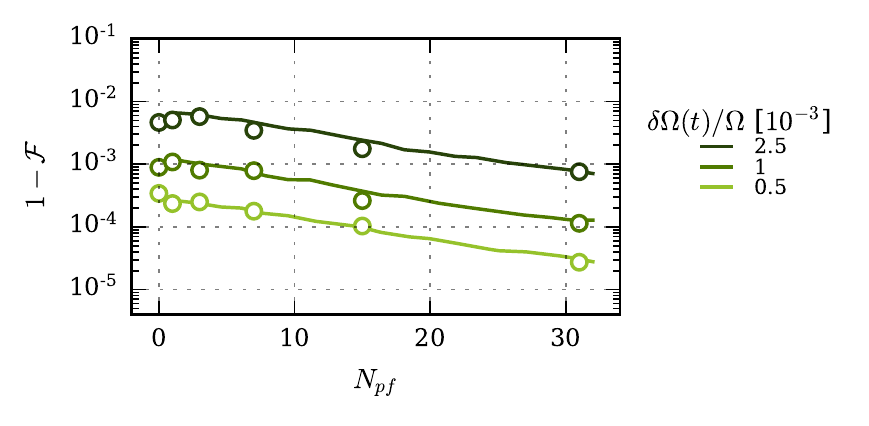}
    \caption{Infidelities of the \msgate\ entangling gate with continuous dynamical decoupling due to amplitude noise in the carrier field itself. A number of phase flips $N_{pf}$ are introduced in the dynamical decoupling field to refocus noise by means of rotary echoes. Circles are the result of numerical simulations (\cf figure \ref{fig:cdd_noise_robustness}) while solid lines are analytical predictions from equation \ref{eq:infidelity_cdd_amp_noise}.}
    \label{fig:cdd_noise_robustness_re}
\end{figure}

While the previous methods of decoupling noise from the drive itself are efficient, they each involve additional fields or modulations which increase the complexity of the scheme. An experimentally simpler solution consists of a rotary echo sequence \cite{solomon1959, gustavsson2012}. Rotary echos are analogous to pulsed dynamical decoupling, however refocussing occurs in a frame rotating with the continuous drive. Instead of introducing $\pi$-pulses, phase flips in the drive are applied throughout the evolution. The filter function of equation \ref{eq:infidelity_cdd_amp_noise} is then found from the timings of the phase flips, and one can use those derived in the context of PDD. The resulting infidelity model is numerically verified in figure \ref{fig:cdd_noise_robustness_re}. We stipulate that mitigating amplitude noise by means of rotary echos is more beneficial than other methods that were mentioned, since phase flips do not affect the power budget and are more straightforward to implement experimentally.

A final infidelity source arises from the off-resonant coupling of the carrier field to the motional states. An order of magnitude estimate of the error is obtained from the excitation probability of the carrier field in the bichromatic interaction picture (see appendix \ref{app:cdd_off_res_coupling}),

\begin{equation} \label{eq:infidelity_cdd_off_res}
\infid = (1 + \frac{\nu^4}{\Omega_c^2\Omega_0^2})^{-1},
\end{equation}
where $\nu$ is the motional frequency and $\Omega_0$ is the sideband Rabi frequency. This interaction is suppressed by ensuring that $\Omega_c \ll \nu$. While large carrier Rabi frequencies are desired to reduce the spin dephasing infidelity term of equation \ref{eq:infidelity_cdd_dephasing}, errors due to off-resonant coupling scale as $\propto \Omega_c^2$ and may quickly deteriorate the fidelity. Nevertheless, Ref. \cite{arrazola2020} shows that the infidelity term of equation \ref{eq:infidelity_cdd_off_res} can be suppressed by introducing a time-dependent phase modulation on all of the fields. The scheme can also be combined with rotary echoes to mitigate imperfections in the carrier drive itself.

\subsubsection{Gate duration}

Since all participating fields are always on, the total available power of the microwave synthesis chain must be shared between several tones: the four MS sidebands and the two dynamical decoupling carriers. This introduces a trade-off in the choice of the carrier Rabi frequency, wherein larger powers are desired to more efficiently decouple the qubit from noise (c.f. equation \ref{eq:infidelity_cdd_dephasing}) but also take away power from the gate fields, thereby reducing the gate speed. Assuming a total power budget $\Omega_{max} = 4 \Omega_{MS} + 2\Omega_c$, the gate duration suffers from a reduction of 

\begin{equation}
\tau_{cdd} = \frac{\Omega_{max}}{\Omega_{max} - 2\Omega_c} \tau_0 
\end{equation}

This result can directly be plugged into the previously derived infidelities of equations \ref{eq:infidelity_cdd_dephasing}, \ref{eq:infidelity_cdd_amp_noise} and \ref{eq:infidelity_cdd_off_res} to find an optimal carrier Rabi frequency $\Omega_c$. Note that this trade-off for CDD is very similar to that of PDD, in that increasing the dynamical decoupling quality by means of increasing the number of $\pi$-pulses or the Rabi frequency leads to an increased gate duration.

\subsubsection{Robustness to static shifts}

\begin{figure}[t!]
    \centering
    \includegraphics[scale=1.]{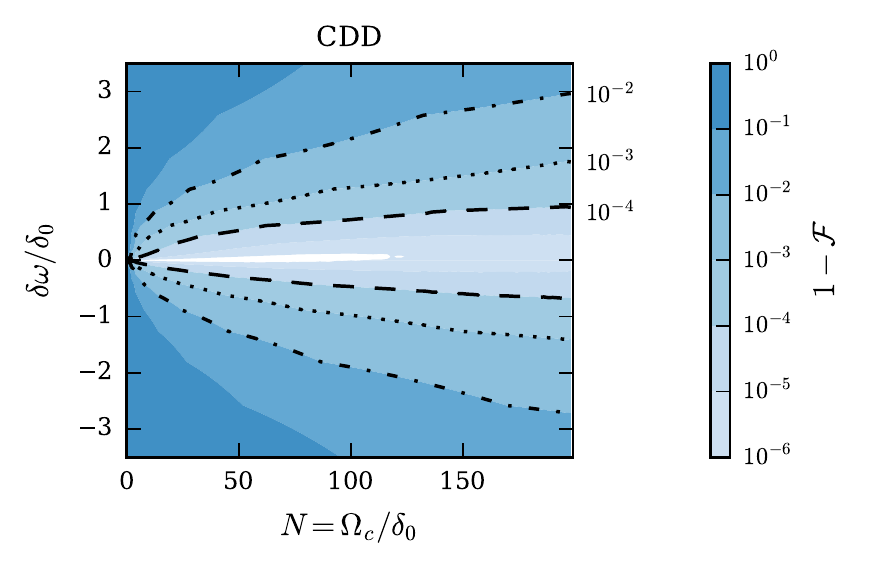}
    \caption{Robustness of the \msgate\ entangling gate protected by continuous dynamical decoupling to static qubit frequency shifts. The Bell state fidelities are numerically simulated for a range of carrier Rabi frequencies $\Omega_c$ and normalised shifts $\delta\omega/\delta_0$, where $\delta_0$ is the MS detuning. The carrier Rabi frequency is parametrized by $N = \Omega_c/\delta_0$, such that $N$ represents the number of 2$\pi$ carrier oscillations completed at the gate duration. The dashed, dotted and dash-dotted lines are contours corresponding to the infidelities $10^{-4}$, $10^{-3}$ and $10^{-2}$.}
    \label{fig:cdd_static_robustness}
\end{figure}

In an identical manner to PDD (\cf section \ref{sec:pdd_static_robustness}), the robustness to static qubit frequency shifts is investigated by numerically simulating the Hamiltonian of equation \ref{eq:hamiltonian_cdd} after replacing $\delta\beta_z(t) \rightarrow \delta\omega$ (see figure \ref{fig:cdd_static_robustness}). The carrier Rabi frequency is parametrized by $N=\Omega_c/\delta_0$, which represents the number of carrier oscillations performed during the interaction. An empirical model is fitted to the resulting infidelities, 

\begin{align} \label{eq:cdd_static_robustness}
\delta\omega/\delta_0 \  \leq 
\begin{cases}
 \ 0.22 \sqrt{N},  \  & \infid \leq 10^{-2},  \\
 \ 0.13 \sqrt{N},  &\infid \leq 10^{-3},  \\
 \ 0.07 \sqrt{N}, & \infid \leq 10^{-4}. 
\end{cases}
\end{align}

This model allows one to estimate the largest tolerable qubit frequency shift that still allows infidelities below a specific threshold. For example, for a gate duration of $\tau_0 = \SI{1}{ms}$, one could achieve infidelities below $10^{-3}$ for shifts of up to $\delta\omega/2\pi = \SI{1.3}{kHz}$ with a carrier Rabi frequency $\Omega_c/2\pi = \SI{100}{kHz}$, as opposed to $\delta\omega/2\pi \approx \SI{1}{kHz}$ with $N_\pi = 100$ pulses using PDD (\cf section \ref{sec:pdd_static_robustness}).

\subsubsection{Calibration requirements}

Similarly to PDD, the CDD scheme requires three fields per ion: two sideband fields driving the MS interaction and one carrier that is continuously applied. Due to the static magnetic field gradient imparting different transition frequencies to each qubit, a total of six fields are required. The calibration DAG is identical to that of the PDD scheme (\cf figure \ref{fig:pdd_cdd_calibrations}); there are 12 nodes, with 10 strong dependencies and 8 weak dependencies.

\subsubsection{Experimental overhead}

The experimental requirements and overhead of the CDD scheme are identical to that of the PDD scheme, and we refer the reader to section \ref{sec:pdd_experimental_req}.

%% file: _sections/Section1/ml_continuous_dd.tex
\subsection{Multi-level Continuous Dynamical Decoupling} \label{sec:ml_continuous_dd}

Both pulsed and continuous dynamical decoupling schemes made use of a simple two-level system and encoded the qubit with a magnetically sensitive transition. Furthermore, both schemes suffer from susceptibility to noise in the dynamical decoupling fields themselves. One can instead make use of the multi-level structure that naturally appears in hyperfine ground states and encode qubits in a decoherence-free subspace. We refer to this approach as \textit{multi-level continuous dynamical decoupling} (MLCDD) \cite{timoney2011, webster2013, aharon2013, mikelsons2015, randall2015, weidt2016, randall2018, webb2018, valahu2021}.  Similarly to CDD, a pair of carrier fields are continuously applied to drive two magnetically sensitive $m_f=\pm 1$ states which we label $\ket{\pm 1}$. The dynamical decoupling Hamiltonian is, for $\sigma_x$ drives,

\begin{equation} \label{eq:mlcdd_hamiltonian_mw}
H_{dd} = \sum_j \frac{\hbar\Omega_c}{2}\left[\frac{1}{\sqrt{2}}\ket{0}^{(j)}\left(\bra{-1}^{(j)} + \bra{+1}^{(j)}\right) + H.C. \right].
\end{equation}

In the eigenbasis of these carriers, one finds the eigenstate $\ket{D} = \frac{1}{\sqrt{2}}(\ket{-1} + \ket{+1})$ whose eigenenergy is degenerate with the $\ket{0'}$ state. For this reason, the transition $\ket{0'}\rightarrow\ket{D}$ is insensitive to noise in the dynamical decoupling drives themselves, and makes for an excellent qubit. To better understand the dynamical decoupling interaction, one can consider noise of the form

\begin{equation} \label{eq:H_noise_mlcdd}
H_{noise} = \frac{\hbar\beta_z(t)}{2} (\ket{+1}\bra{+1} - \ket{-1}\bra{-1}),
\end{equation}

where we only include the first order Zeeman shifts of the magnetically sensitive transitions, as this generally dominates the infidelity. Transforming $H_{noise}$ in an interaction picture with respect to $H_{dd}$, $e^{-itH_{dd}/\hbar}H_{noise}e^{itH_{dd}/\hbar}$,

\begin{equation} \label{eq:cdd_hnoise_int_pic}
\tilde{H}_{noise} = \frac{\delta\beta_z(t)}{2\sqrt{2}} \left( S_+ e^{it\Omega_c/\sqrt{2}} + H.C. \right).
\end{equation}

Here, the multi-level ladder operators $S_+ = \ket{D}\bra{d} + \ket{u}\bra{D}$ and $S_- = \ket{d}\bra{D} + \ket{D}\bra{u}$ describe transitions between the eigenstates. The effects of the dynamical decoupling drive are elucidated by equation \ref{eq:cdd_hnoise_int_pic}, which shows that only noise nearly resonant with the eigenstate splitting $\Omega_c/\sqrt{2}$ may drive population out of the logical qubit state $\ket{D}$ and into spectator states $\ket{u}$ and $\ket{d}$. Consequently, dephasing of the magnetic sensitive states under the action of continuous dynamical decoupling results in an equal distribution of population among the eigenstates, and a third of the population remains in $\ket{D}$ at large durations. In the interaction picture with respect to the continuous drives, dephasing therefore becomes analogous to a leakage error mechanism. Finally, we note that transitions within the eigenstates are slightly weaker by a factor of $1/\sqrt{2}$.

Obtaining an MS type interaction within the dressed states involves applying RF fields on either of the $\ket{0'}\rightarrow\ket{\pm 1}$ transitions, and introducing a detuning equal to the motional frequency. In this way, sideband transitions are driven on the $\ket{0'}\rightarrow\ket{D}$ transition. Note that, due to the second order Zeeman shift, both $\ket{0'}\rightarrow\ket{\pm 1}$ transitions are resolvable and can lead to transitions with the $\ket{D}$ state \cite{webster2013}. The full Hamiltonian describing the bichromatic interaction is 

\begin{align} \label{eq:mlcdd_hamiltonian_rf}
H_{rf} = \  \sum_j  \frac{\hbar\Omega_{rf}}{2} \big( \ket{+1}^{(j)}\bra{0'}^{(j)} e^{-i\delta_{rf}t} e^{\epsilon_j (a^\dagger - a)} \nonumber \\
+   \ket{-1}^{(j)}\bra{0'}^{(j)} e^{i(\delta_{rf} -\Delta\omega_\pm)t} e^{-\epsilon_j(a\dag - a)} + \textrm{H.C.} \big),
\end{align}

where $\delta_{rf}$ is the detuning from the carrier transition, $\epsilon_j$ is the Lamb-Dicke parameter and $\Delta\omega_{\pm}$ is the frequency splitting between the $\ket{0'}\rightarrow\ket{+1}$ transitions that arises due to the second order Zeeman shift. Moving $H_{rf}$ into an interaction picture with respect to $H_{dd}$ of equation \ref{eq:mlcdd_hamiltonian_mw}, and dropping fast oscillating terms under a rotating wave approximation in the limits $\epsilon \Omega_{rf}/2 \ll \Delta\omega_\pm$ and $\delta_0 \ll \Omega_c/\sqrt{2} \ll \nu$ (where $\delta_0 = \delta_{rf} - \nu$), 

\begin{equation}
\tilde{H}_{rf} = - \frac{\epsilon\Omega_0}{2} (\tilde{\sigma}_x^{(1)} - \tilde{\sigma}_x^{(2)} ) \left(a e^{i \delta_0 t} - a^\dagger e^{-i\delta_0 t} \right).
\end{equation}

This final Hamiltonian corresponds to the usual \msgate\ interaction, where $\tilde{\sigma}_x^{(i)} = \ket{0'}\bra{D} - \ket{D}\bra{0'}$ is the modified Pauli operator, and $\Omega_0 = \Omega_{rf}/\sqrt{2}$ is the MS Rabi frequency.

\begin{figure}[t!]
    \centering
    \includegraphics[scale=1.]{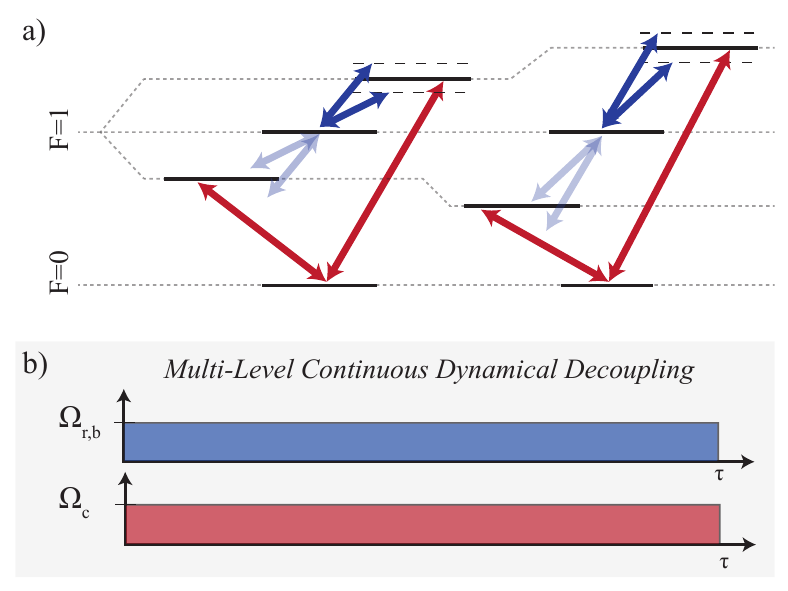}
    \caption{Required fields (a) and pulse sequence (b) for a MS entangling gate that uses the multi-level continuous dynamical decoupling scheme. The RF gate fields (blue) are near resonance with a magnetic sensitive transition within the $F=1$ manifold. The microwave dynamical decoupling fields (red) bridge both magnetic sensitive transitions between $F=0$ and $F=1$.}
    \label{fig:ml_cdd_schematic}
\end{figure}

\subsubsection{Fidelity}

The decay function of the spin's coherence under MLCDD is slightly different than for PDD and CDD (\cf equation \ref{eq:infidelity_spin_dephasing}). As outlined in the noise mechanism of equation \ref{eq:cdd_hnoise_int_pic}, dephasing causes population leakage into spectator eigenstates. In the limit of large durations, the state decoheres into an equal distribution of $\ket{D}$, $\ket{u}$ and $\ket{d}$. Taking this into account, the modified infidelity from equation \ref{eq:infidelity_spin_dephasing} becomes

\begin{equation}
\infid = \frac{1}{3}(1 - e^{-\chi(t)} ).
\end{equation}

The filter function of the MLCDD scheme is almost identical to that of CDD, \ie the transfer function corresponds to a sinc-like passband filter that can be approximated by a Dirac delta function in the limit of small bandwidths (c.f. appendix \ref{app:filter_functions}). Conversely, the passband filter's center frequency is $\Omega_c/\sqrt{2}$ instead of $\Omega_c$ due to the different eigenenergies of the dressed states. With these modifications, the infidelity due to dephasing becomes, to first order,

\begin{equation} \label{eq:mlcdd_theory_infidelity}
\infid = \frac{S_z(\Omega_c/\sqrt{2})\tau}{12}.
\end{equation}

The interaction is protected from noise in the continuous dynamical decoupling fields themselves, since the transition frequency of the qubit $\{\ket{0'}, \ket{D}\}$ is independent of the amplitude $\Omega_c$. Therefore, the MLCDD scheme does not have infidelity terms originating from the drive's noise. Nevertheless, the multiple levels that are used in this scheme lead to several unwanted transitions and parasitic couplings.

\begin{figure}[t!]
	\center
    \includegraphics[scale=1.05]{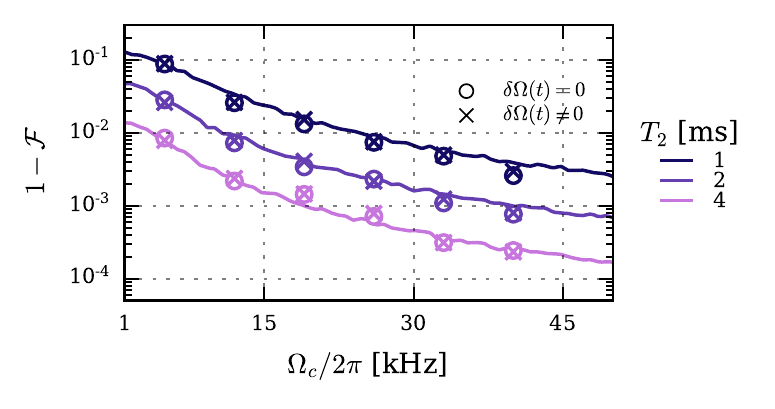}
    \caption{Numerical simulations of the \msgate\ entangling gate protected by multi-level continuous dynamical decoupling and subject to spin dephasing noise. See caption of figure \ref{fig:pdd_noise_robustness} for numerical simulation details, namely, the Hamiltonians of equations \ref{eq:mlcdd_hamiltonian_mw}, \ref{eq:H_noise_mlcdd} and \ref{eq:mlcdd_hamiltonian_rf} were used. Analytical fidelities were obtained from equation \ref{eq:mlcdd_theory_infidelity}. Numerical simulations modelled both an ideal (circles) and a noisy (crosses) dynamical decoupling carrier.}
    \label{fig:mlcdd_noise_robustness}
\end{figure}

A dominant infidelity term of order $\Omega_c^2/\nu^2$ appears from the off-resonant coupling of the carrier fields to the motional sidebands \cite{mikelsons2015}. This term, however, oscillates at a frequency $\nu$, and can therefore be eliminated by carefully choosing the gate time. This nevertheless requires a timing resolution that is much smaller than $1/\nu$, and the interaction is more susceptible to drifts of the motional mode frequency. Another similar error term of order $\Omega_0^2/\nu^2$ originates from the off-resonant coupling of the MS fields to the carrier transitions. Nevertheless, amplitude pulse shaping can be used to adiabatically drive transitions and reduce the timing sensitivity at the gate duration \cite{weidt2016}. 

Other higher order terms arise from the coupling of the sideband fields to the carrier transitions, leading to population leakage and unwanted energy shifts \cite{valahu2021}. These terms can be made vanishingly small by driving the carrier transitions between the two qubits at different Rabi frequencies. 

Finally, off-resonant coupling of the carriers to the motional sidebands as well as coupling of the sideband fields to the spectator's motional states generate an MS type interaction of strength $\Omega_c^2/(2\nu^2 - \Omega_c^2)$. This interaction can nevertheless be compensated for by appropriately adjusting the gate's duration.

\subsubsection{Gate duration}

The gate speed of the MLCDD scheme is different from the PDD and CDD methods in that the gate fields (RF) and dynamical decoupling fields (microwave) originate from separate synthesis chains. Their power budgets are therefore independent, and there is no longer a trade-off between increasing the dynamical decoupling quality and reducing the interaction strength. The gate duration is then

\begin{equation}
\tau_{MLCDD} = \tau_0,
\end{equation}

where $\tau_0$ is the duration of a primitive MS entangling gate after replacing the usual Rabi frequency with $\Omega_0 \rightarrow \Omega_{rf}$.

\subsubsection{Robustness to static shifts} \label{sec:mlcdd_robustness_static}

Previously, the robustness of the PDD and CDD schemes were investigated by numerically simulating a static shift in the carrier transition. This reflects a robustness to both a magnetic field shift, as well as an imperfect frequency in the control fields set by the experimentalist. In the MLCDD scheme, however, these two error sources do not coincide with the same robustness. On the one hand, a change in the magnetic field will, to first order shift the magnetic sensitive transitions by an equal and opposite amount (\cf equation \ref{eq:H_noise_mlcdd}). On the other hand, a misset in the frequency of the control field results in a detuning error with the $\ket{0'}\rightarrow\ket{D}$ transition. The robustness to either, as will be shown, is very different.

\begin{figure}[t!]
    \centering
    \includegraphics[scale=1.]{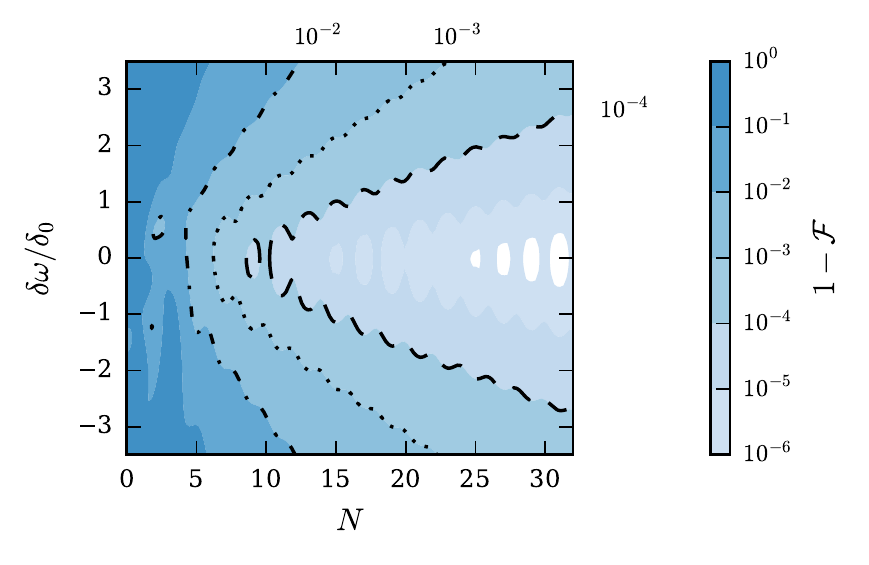}
    \caption{Robustness of the \msgate\ entangling gate protected by multi-level continuous dynamical decoupling to static qubit frequency shifts. The Bell state fidelities are numerically simulated for a range of carrier Rabi frequencies $\Omega_c$ and normalised shifts $\delta\omega/\delta_0$, where $\delta_0$ is the MS detuning. The carrier Rabi frequency is parametrized by $N = \Omega_c/\delta_0$, such that $N$ represents the number of 2$\pi$ carrier oscillations completed at the gate duration. The dashed, dotted and dash-dotted lines are contours corresponding to the infidelities $10^{-4}$, $10^{-3}$ and $10^{-2}$.}
    \label{fig:ml_cdd_static_robustness}
\end{figure}

We first consider the robustness to a change in magnetic field by numerically simulating the Hamiltonians of equations \ref{eq:mlcdd_hamiltonian_mw}, \ref{eq:H_noise_mlcdd} and \ref{eq:mlcdd_hamiltonian_rf} (after dropping terms rotating with $\Delta\omega_{\pm 1}$ and $\nu$) and performing the replacement $\beta_z(t) \rightarrow \delta\omega$. The results are reported in figure \ref{fig:ml_cdd_static_robustness} for a range of carrier Rabi frequencies, where $N=\Omega_c/\delta_0$. An empirical model is then constructed by fitting the linear portion of the contours ($N \gg 1$),

\begin{align} \label{eq:ml_cdd_static_robustness}
\delta\omega/\delta_0 \  \leq 
\begin{cases}
 \ -0.33 +  0.30N,  \  & \infid \leq 10^{-2},  \\
 \ -0.40 + 0.17N,  &\infid \leq 10^{-3},  \\
 \ -0.53 + 0.10 N, & \infid \leq 10^{-4}. 
\end{cases}
\end{align}

The model of equation \ref{eq:ml_cdd_static_robustness} suggests that the carrier dynamical decoupling fields of the MLCDD are very efficient at mitigating errors due to shifts in the magnetic field.

We now turn our attention to shifts in the control fields, which is simulated by replacing the noise Hamiltonian of equation \ref{eq:H_noise_mlcdd} with $H_{noise} = \delta\omega/2 (\ket{D}\bra{D} - \ket{0'}\bra{0'})$. Note that a shift of this nature could also occur from stark shifts due to the sideband fields \cite{weidt2016, webb2018}. With this new definition of $H_{noise}$, we effectively retrieve the robustness of a primitive \msgate\ gate within a two-level system in the absence of any dynamical decoupling,

\begin{align} \label{eq:ml_cdd_static_robustness_D}
\delta\omega/\delta_0 \  \leq 
\begin{cases}
 \ \num{2.8e-2},  \  & \infid \leq 10^{-2},  \\
 \ \num{0.9e-2},  &\infid \leq 10^{-3},  \\
 \ \num{0.3e-2}, & \infid \leq 10^{-4}. 
\end{cases}
\end{align}

The robustness demonstrated by equation \ref{eq:ml_cdd_static_robustness_D} indicates that the MLCDD scheme is much less forgiving to static errors arising from imperfect experimental knowledge, \eg calibrations that aren't precise enough or unaccounted stark shifts. Furthermore, infidelities from these errors can not be reduced by increasing the dynamical decoupling power. 

\subsubsection{Calibration requirements}

\begin{figure}[t!]
    \centering
    \includegraphics[scale=1.]{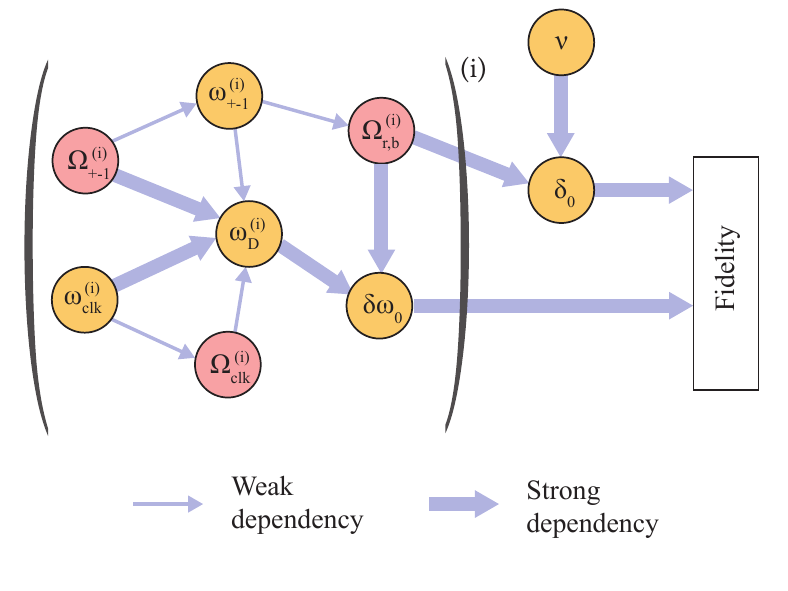}
    \caption{Directed acyclic graph describing the calibration requirements of the multi-level continuous dynamical decoupling scheme. Labels and caption identical to figure \ref{fig:pdd_cdd_calibrations}.}
    \label{fig:ml_cdd_calibrations}
\end{figure}

The calibration requirements of the MLCDD scheme are determined from the corresponding Directed Acyclic Graph (DAG) (\cf section \ref{sec:pdd_calibration_requirements}). Four fields are required per ion: a pair of sideband fields which drive the MS interaction and a pair of carrier dynamical decoupling fields. The DAG, illustrated in figure \ref{fig:ml_cdd_calibrations}, contains nodes corresponding to the frequencies and amplitudes of each of these fields. Additional nodes are included to calibrate the $\ket{0}\rightarrow\ket{0'}$ clock transition ($\omega_{clk}$ and $\Omega_{clk}$), since it is required to map in and out of the decoherence free subspace \cite{randall2018}. Each ion requires the calibrations of the $\ket{0}\rightarrow\ket{\pm1}$ transition frequencies $\omega_{\pm 1}$ and amplitudes $\Omega_{\pm 1}$, as well as the blue and red sideband Rabi rates $\Omega_{r,b}$. One can then calibrate the qubit transition frequency $\omega_D$ and measure the stark shift $\delta\omega_0$. The final nodes in the DAG are the secular frequency $\nu$ and the MS detuning $\delta_0$. In total, the DAG comprises of 22 parameters, with 20 strong dependencies and 14 weak dependencies.

\subsubsection{Experimental overhead}

The MLCDD scheme requires 2 MS sideband fields and 2 dynamical decoupling drives per ion for a total of 8 fields. The carrier drives should be resonant with the hyperfine $\ket{F=0, m_f=0}\rightarrow\ket{F=1,m_f=\pm 1}$ transitions, requiring near equal amounts of $\sigma^+$ and $\sigma^-$ polarization. The amplitudes of the carrier fields within an ion should be identical in order to maximize the coherence of the eigenstate $\ket{D}$ \cite{randall2016}. The sideband fields should be detuned from either of the $\ket{F=1,m_f=0}\rightarrow\ket{F=1,m_f=\pm 1}$ transitions within the $\ket{F=1}$ triplet, requiring a single component of $\sigma^+$ or $\sigma^-$ polarization. 

An additional microwave field resonant with the $\ket{F=0,m_f=0}\rightarrow\ket{F=1,m_f=0}$ clock transition is required to map population in and out of the computational subspace $\{\ket{0'}, \ket{D}\}$ \cite{randall2015, randall2018}, necessitating a component of $\pi$ polarization.

%% file: _sections/Section1/summary.tex
\subsection{Summary}\label{sec:spin_robustness_summary}

\input{_sections/Section1/summary_table.tex}

The performances of PDD, CDD and MLCDD are summarized in table \ref{tab:gate_summary}. We first note that the resilience of all three schemes to spin dephasing is very similar. Within the filter function framework, the transfer functions that are applied to the noise's PSD are akin to a narrow bandwidth passband filter. The filter's center frequency is proportional to the number of $\pi$-pulses for PDD and the carrier Rabi frequency for CDD and MLCDD. Increasing these parameters is desirable to displace the bandpass filter to higher frequencies which, for a typical correlated noise spectrum, contains lower powers of noise. The filter functions of the pulsed and continuous schemes differ slightly in that CDD and PDD implement a sinc-like function that can be approximated by a Dirac delta, while the PDD implements a high-pass filter, whose low-frequency roll-off is highly dependent on the timings of the $\pi$ pulses. 

The PDD and CDD schemes have an additional source of infidelity that arises from the dynamical decoupling fields themselves. In the pulsed scheme, noise and imperfections in the $\pi$-pulses can be treated as static errors that accumulate with the number of pulses. For the continuous scheme, amplitude noise in the carrier may decohere the spins and increases with the carrier Rabi frequency. In both cases, this introduces a trade-off with the number of pulses (PDD) and the carrier Rabi frequency (CDD). Nevertheless, one can implement noise mitigation extensions such as pulses along alternating axes for PDD and rotary echoes for CDD. In the case of the MLCDD scheme, errors in the continuous drives do not lead to infidelities by design. Furthermore, leading terms in the full Hamiltonian suggest that larger carrier Rabi frequencies affect the interaction by introducing static qubit frequency shifts and modified interaction strengths, both of which can be corrected for by calibrations. We therefore do not include an intrinsic infidelity term, however precautions should be taken as higher order terms that are analytically and numerically difficult to analyse may lead to infidelities in certain parameter regimes. 

The gate durations of the PDD and CDD scheme are subject to a similar trade-off, in that increasing the quality of the dynamical decoupling leads to prolonged gate durations. Along with the intrinsic infidelities appearing from noise in the dynamical decoupling fields themselves, this offers a more practical optimisation constraint. This additional trade-off does not appear within the MLCDD scheme, since the gate duration depends only on the MS field's power which originates from a different physical signal source to the carrier dynamical decoupling fields. 

The calibration requirements of the PDD and CDD scheme are much lower than for MLCDD. This can be attributed to the MLCDD schemes using multiple levels within the hyperfine ground state to encode a qubit, necessitating many more calibrations for each transition. PDD and CDD however make use of an approximate 2-level system, greatly reducing the number of parameters to be calibrated. A similar comment can be made on the experimental implementation and overhead. The added benefits of MLCDD come at the cost of a greater experimental complexity. For example, one requires a synthesis chain for both RF and microwaves, all types of polarization and stringent requirements on the amplitudes between carrier fields.

From the results of table \ref{tab:gate_summary} and from previous discussions, it has hopefully become clear to the reader that there is not one quantum control method which solves every experimental problem, be it high fidelities or fast gates. On the contrary, each gate scheme addresses a particular subset of a problem, while introducing a new set of constraints and errors. It is therefore important to characterize the various trade-offs to understand the suitability of one scheme over another. 

%% file: _sections/Section1/summary_table.tex
\setlength{\tabcolsep}{0.5em} 
{\renewcommand{\arraystretch}{1.5}
\newcolumntype{P}[1]{>{\centering\arraybackslash}p{#1}}

\begin{table*}[t!]
\begin{tabular}{P{0.13\textwidth}P{0.04\textwidth}P{0.2\textwidth}P{0.15\textwidth}P{0.12\textwidth}P{0.15\textwidth}}
\hline\hline
\textbf{} & & \textbf{PDD} & \textbf{CDD} & \multicolumn{2}{c}{\textbf{MLCDD}}\\ \hline
\multicolumn{2}{l}{Infidelity:} & & & \multicolumn{2}{c}{} \\
\multicolumn{2}{l}{\hspace{3mm}\textit{Dephasing}} &   $\frac{1}{2}(1 - e^{-\chi(\tau,N_\pi)})$   &  $\frac{S_z(\Omega_c)\tau}{4}$  &  \multicolumn{2}{c}{$\frac{S_z(\Omega_c/\sqrt{2}\tau}{12}$} \\
\multicolumn{2}{l}{\hspace{3mm}\textit{Intrinsic}}  &  $\frac{Tr(UU_0)}{Tr(U)Tr(U_0)}$  & $\frac{\Omega_c^2}{2\pi}\int^{+\infty}_{-\infty}d\omega S_x(\omega) \frac{F_x(\omega,\tau)}{\omega^2}$ & \multicolumn{2}{c}{Higher order terms} \\ 
\multicolumn{2}{l}{} &     & $(1 + \frac{\nu^4}{\Omega_c^2\Omega_0^2})^{-1}$ &  \multicolumn{2}{c}{} \B \\ \hline
\multicolumn{2}{l}{Gate time :}   & $\tau_0 + N_\pi \tau_\pi$  & $\frac{\Omega_{max}}{\Omega_{max} - 2\Omega_c}\tau_0,$ & \multicolumn{2}{c}{$\tau_0$ (RF)} \\ \hline
Robustness [\%]: &  &  &  & \textit{B-field shift} &  \textit{Control field shift} \\\cline{5-6}
\multicolumn{2}{l}{\hspace{6mm}$10^{-2}$}  & $(2.8 + 3.2 N_\pi)\times 10^{-2}$ & $0.22 \sqrt{N}$ & $-0.33 + 0.3 N$ & $\num{2.8e-2}$  \\
\multicolumn{2}{l}{\hspace{6mm}$10^{-3}$} & $(0.8 + N_\pi)\times 10^{-2}$   & $0.13 \sqrt{N}$ & $-0.40 + 0.17 N$   & $\num{0.9e-2}$  \\ 
\multicolumn{2}{l}{\hspace{6mm}$10^{-4}$} & $(0.3 + 0.3 N_\pi)\times 10^{-2}$ & $0.07 \sqrt{N}$  &  $-0.53 + 0.10 N$   & $\num{0.3e-2}$  \\ \hline
\multicolumn{2}{l}{Calibrations :} & \{12, 12, 10\} & \{12, 12, 10\}  & \multicolumn{2}{c}{\{22, 20, 14\}} \\  \hline
  \multicolumn{2}{l}{Experimental Req.} & & & & \\
  \multicolumn{2}{l}{\hspace{3mm}\textit{Number of fields}} & \multicolumn{1}{l}{ 6$\times$MW} & \multicolumn{1}{l}{ 6$\times$MW} & \multicolumn{2}{l}{ 4$\times$MW and 4$\times$RF} \\
  \multicolumn{2}{l}{\hspace{3mm}\textit{Bandwidth}} & \multicolumn{1}{l}{10 MHz} & \multicolumn{1}{l}{10 MHz} & \multicolumn{2}{l}{MW: 50 MHz } \\
 & & & & \multicolumn{2}{l}{RF:  10 MHz } \\
 \multicolumn{2}{l}{\hspace{3mm}\textit{Polarisation}} & \multicolumn{1}{l}{\{$\pi$, $\sigma^+$\} or \{$\pi$, $\sigma^-$\}} & \multicolumn{1}{l}{\{$\pi$, $\sigma^+$\} or \{$\pi$, $\sigma^-$\}} & \multicolumn{2}{l}{MW: \{$\pi$, $\sigma^+$, $\sigma^-$\}} \\
  & & &  & \multicolumn{2}{l}{RF: \{$\sigma^+$\} or \{$\sigma^-$\}} \\
 \hline\hline
\end{tabular}
\caption{\label{tab:gate_summary} Summary of the performances of pulsed dynamical decoupling (PDD), continuous dynamical decoupling (CDD) and multi-level continuous dynamical decoupling (MLCDD). The various metrics are obtained from sections \ref{sec:pulsed_dd}, \ref{sec:continuous_dd} and \ref{sec:ml_continuous_dd}. The symbols and nomenclature are explained throughout the main text. The infidelities are classified into dephasing errors, representing dephasing noise that is mitigated by the dynamical decoupling scheme, and intrinsic errors, which consider noise arising from the scheme itself. The MLCDD scheme contains an empirical robustness model for shifts arising from magnetic field changes, and for shifts due to errors in the control field's frequency (\cf section \ref{sec:mlcdd_robustness_static}). The calibration triplets represent \{\textit{number of parameters, weak dependencies, strong dependencies}\}.}
\end{table*}

%% file: _sections/Section2/section2.tex
\section{Robustness to motional decoherence} \label{sec:robustness_motional_decoherence}

\input{_sections/Section2/introduction.tex}

\input{_sections/Section2/pst_framework.tex}

\input{_sections/Section2/pst_engineering.tex}

\input{_sections/Section2/comparison.tex}

\input{_sections/Section2/optimal_pst.tex}

\input{_sections/Section2/thermal_noise.tex}

\input{_sections/Section2/summary_and_outlook.tex}

%% file: _sections/Section2/introduction.tex
Motional decoherence encapsulates all mechanisms that decohere the common modes of motion of a trapped ion chain. Since the \msgate\ type entangling gates presented in this manuscript use the motion as an information bus, motional decoherence will inherently reduce their overall fidelity. We identify three motional decoherence mechanisms: \textit{motional heating}, \textit{motional dephasing} and \textit{thermal noise}. \textit{Motional heating} involves a phonon gain and is caused by the ion chain coupling to its environment. \textit{Motional dephasing} is analogous to spin dephasing and describes frequency fluctuations of the vibrational modes. Finally, \textit{thermal noise} arises from the intrinsic noise of ions in a thermal state. Note that thermal noise in itself is not a source of infidelity, as the Mølmer-Sørensen interaction is by design independent of the initial thermal state of the ions. However, large temperatures increase the sensitivity of the fidelity to various other parameters and inevitably result in a loss of fidelity. 

There exists a variety of schemes that add robustness to motional decoherence. The literature contains a vast amount of work that theoretically and experimentally investigate these schemes. It is not immediately clear, however, as to how their motional dynamics differ from one another. In this section, we consolidate the breadth of work pertaining to motional robustness by introducing a unifying theoretical framework. This allows us to clearly understand the advantages of certain implementations and choose an optimal robust scheme.

%% file: _sections/Section2/pst_framework.tex
\subsection{Robust phase space trajectories} \label{sec:robust_psts}

Achieving robustness to motional decoherence amounts to engineering efficient trajectories of the ion's motion in phase space. The Phase Space Trajectory (PST) has a start, an end and a time-dependent position. Note that the phase space referred to here is similar to that of a simple harmonic oscillator, for which the two-dimensional space is defined by the position and momentum. We will show that the robustness of a scheme is entirely dependent on the path that is taken by the PST. In this way, increasing the resilience of a scheme to motional decoherence can be thought of as a geometric optimisation problem, since it involves optimising a constrained 2-dimensional trajectory.

The PST of the $k^{th}$ motional mode that arises from a MS interaction with a pair of ions is

\begin{equation} \label{eq:generalized_pst}
\alpha_{k}(t) = \epsilon_k \int_0^t dt' \Omega(t')e^{i\delta_k(t') t}e^{-i\phi(t')},
\end{equation}

where $\epsilon_k$ is the Lamb-Dicke parameter. The Rabi frequency, phase, and frequency are assumed to be equal for all ions participating in the entangling operation (a valid assumption for global microwave fields). If all parameters are time-independent, equation \ref{eq:generalized_pst} becomes $\alpha_k(t) = \frac{i\epsilon\Omega}{2\delta}e^{-i \phi}(e^{i\delta t} - 1) $, which is the trajectory of a circle. For an arbitrary set of parameters, the infidelity from imperfection in the PSTs is \cite{bentley2020}

\begin{widetext}
\begin{equation} \label{eq:off_res_coupling_spectator_mode}
\infid = 1- \left| \left(  \prod_m \prod_n \cos(\Phi_{m,n}- \Psi_{m,n}) \right) \left( 1 - \sum_k \sum_j \left[ |\alpha_{k}(\tau)|^2 (\nbar_k + \frac{1}{2}) \right] \right) \right|^2,
\end{equation}
\end{widetext}

where $\nbar_k$ is the average motional occupation. The first factor multiplies over all pairs of ions $(m,n)$ and evaluates errors between the desired entangling phase $\Phi_{m,n}$ and the obtained phase $\Psi_{m,n}$. Errors in $\Psi_{m,n}$ arise from deviations in the enclosed area of the target mode, or parasitic phase accumulations of the spectator motional modes. The second factor of equation \ref{eq:off_res_coupling_spectator_mode} captures infidelities from residual spin-motion coupling for all modes $k$ and ions $j$. If $\alpha_{k}(\tau) = 0$, this second infidelity term reduces to zero, regardless of the mode's temperature $\nbar_k$. This further demonstrates the \msgate\ gate's robustness to the ion chain's temperature. However, if the PST does not finish at the origin ($\alpha_{k}(\tau) \neq 0$), larger $\nbar_k$ amplify the infidelity term. This places a practical constraint on the maximal temperature, since experimentally achieved PSTs are never ideal. Furthermore, this leads to the first and most fundamental requirement for the PSTs: the paths from all participating motional modes must return to the origin at the gate time $\tau$,

\begin{equation} \label{eq:pst_requirement_finish_center}
\mathcal{R}_{end} \coloneqq \sum_k |\alpha_k(\tau)|^2 = 0.
\end{equation}

In the primitive \msgate\ interaction, this is ensured by carefully choosing the gate's parameters, \ie setting $\delta_0 = 2\epsilon\Omega_0$ and $\tau = 2\pi/\delta_0$.

\subsubsection{Motional dephasing}

We now turn our attention towards robustness to motional dephasing, and first focus on time-independent static frequency shifts. This is modelled by adding a small error to the detuning in equation \ref{eq:generalized_pst}, $\delta_k \rightarrow \delta_k + \delta$. The PST's sensitivity to motional frequency shifts is eliminated to first order by setting $\partial \alpha_k(\tau)/\partial\delta = 0$, and one finds \cite{leung2018, bentley2020}

\begin{equation}
i\tau \alpha_k(\tau) - i \int^\tau_0 dt \alpha_k(t) = 0.
\end{equation}

The $\alpha(\tau)$ of the first term evaluates to zero as it is fulfilled by the requirement $\mathcal{R}_{end}$ of equation \ref{eq:pst_requirement_finish_center}. The second term describes the average position of the PST throughout the evolution. The PST should therefore be designed such that this term reduces to zero. The requirement for robustness to static motional frequency offsets of the $k^{th}$ mode is then

\begin{equation} \label{eq:pst_requirement_average_center}
\mathcal{R}_{dephasing} \coloneqq \alpha_{av} = \int^\tau_0 dt \alpha_k(t) = 0.
\end{equation}

The primitive \msgate\ interaction does not satisfy $\mathcal{R}_{dephasing}$ as the average position described by the circular PST is $\epsilon\Omega/2\delta$. This explains why the primitive gate is sensitive to static and time-dependent motional frequency noise to first order. It is also interesting to note that the requirements of equations \ref{eq:pst_requirement_finish_center} and \ref{eq:pst_requirement_average_center} can be simplified by only considering PSTs that are symmetric about an axis. If this is the case, fulfilling \ref{eq:pst_requirement_average_center} will by definition satisfy \ref{eq:pst_requirement_finish_center}. 

Having provided a constraint for robustness to static errors, we now show robustness to time-dependent fluctuations of the motional frequency. We take a similar approach to \ref{sec:robustness_spin_dephasing} and provide an expression for the infidelity within the filter function framework. The noise under consideration can be modelled as $\beta_\delta(t) a\dag a$ fluctuations with PSD $S_\delta(\omega)$. The infidelity is \cite{milne2020}

\begin{equation}
\mathcal{I} = \frac{1}{2\pi} \int^{+\infty}_{-\infty} d\omega S_\delta(\omega) F_\delta(\omega).
\end{equation}

The filter function is summed over all motional modes, $F_\delta(\omega) = \sum_k F_{\delta, k}(\omega)$. For the $k^{th}$ mode, the filter function corresponding to a pair of ions $m,n$ is

\begin{align} \label{eq:filter_function_motional_dephasing}
& F_{\delta,k}(\omega) = \nonumber \\
& \frac{T_k}{4} (|\epsilon_k^{(m)}|^2 + |\epsilon_k^{(n)}|^2)|\int^{\tau}_0 dt \Omega(t) e^{i(\delta_k(t)-\omega) t}e^{-i \phi(t)} t|^2,
\end{align}
where we have defined $T_k = 2 (\nbar_k + \frac{1}{2})$. Satisfying the robustness condition of \ref{eq:pst_requirement_average_center} will minimize the integral term of equation \ref{eq:filter_function_motional_dephasing} for small $\omega$, thereby increasing the efficiency of the filter function.

\subsubsection{Motional heating}

We here consider robustness to motional heating. In phase space, one can think of heating as random kicks which disturb the PST, preventing it from returning to the origin. We can therefore put forth a qualitative requirement and stipulate that the average distance from the origin must be minimized to mitigate the effects of motional heating. Quantitatively, the distance of the PST at any time $t$ is $ |\alpha_k(t)|^2$, and the time-averaged distance is $\langle|\alpha_k(t)|^2\rangle = \frac{1}{\tau}\int^{\tau}_0 dt|\alpha_k(t)^2|$. In \myref \cite{haddadfarshi2016}, this is placed on a firmer footing and it is found that the best performance is obtained when $\langle|\alpha_k(t)|^2\rangle$ is minimized. This allows us to write the final requirement for robustness to motional heating, 

\begin{equation}\label{eq:pst_requirement_min_heating}
\mathcal{R}_{heating} \coloneqq \textrm{min}\left(\frac{1}{\tau}\int^\tau_0dt |\alpha_k(t)|^2 \right).
\end{equation}

The average distance can be related to the heating rate in order to provide a more useful metric of comparison. The system is first described with a master equation, $d\rho/dt = -i[H, \rho] + \mathcal{L}(\rho)$, where $\mathcal{L}(\rho)$ is a Lindbladian operator. Assuming the usual Lindblad heating operators \cite{sorensen2000, haddadfarshi2016} with a heating rate $\ndot$, the reduced density matrix becomes

\begin{equation}
\frac{d}{dt}\rho_{M_y, M_y'} = -(M_y - M_y')^2 \ndot |\alpha(t)|^2 \rho_{M_y, M_y'},
\end{equation}
and after integrating over the gate duration and solving for $\rho$,

\begin{equation} \label{eq:rho_heating_evolution}
\rho_{M_y, M_y'}(\tau) = \rho_0 \textrm{Exp}[-(M_y - M_y')^2\ndot \int^{\tau}_0dt|\alpha_k(t)|^2].
\end{equation}

Here, $M_y$ and $M_y'$ are eigenvalues of the \msgate\ spin operators. The infidelity associated with creating a maximally entangled Bell state is derived from equation \ref{eq:rho_heating_evolution}, and after setting $\int^{\tau}_0dt|\alpha_k(t)|^2 = \langle |\alpha_k|^2\rangle \tau$,

\begin{equation} \label{eq:infidelity_heating}
\mathcal{I} = \frac{5}{8} - \frac{1}{2} \textrm{Exp}[- \ndot \langle |\alpha_k|^2\rangle \tau ]- \frac{1}{8} \textrm{Exp}[-4 \ndot \langle|\alpha_k|^2\rangle \tau ].
\end{equation}

Taking the first order Taylor expansion of equation \ref{eq:infidelity_heating}, the infidelity due to motional heating is approximated by

\begin{equation} \label{eq:infidelity_heating_simple}
\infid = \ndot \langle |\alpha_k|^2\rangle \tau.
\end{equation}

From equation \ref{eq:infidelity_heating_simple}, it becomes clear that minimizing the average distance from the origin $ \langle |\alpha_k|^2\rangle$ will minimize the infidelity. For a primitive two-qubit \msgate\ gate with constant parameters, the average displacement is $\langle |\alpha_k|^2\rangle_0 = 1/2$. The infidelity therefore simplifies to $\infid = \ndot \tau / 2$. Alternatively, the heating rate may be replaced with an effective heating rate such that

\begin{equation} \label{eq:reduction_factor_heating}
\ndot_{eff} = \ndot R_{heat}, \ \ \ R_{heat} = \frac{\langle |\alpha_k|^2\rangle}{\langle |\alpha_k|^2\rangle_0},
\end{equation}
and the infidelity becomes

\begin{equation}\label{eq:reduction_factor_heating_simple}
\infid = \frac{1}{2}\ndot_{eff} \tau.
\end{equation}

Schemes which mitigate motional heating can therefore be thought of as providing improvements to the heating rate through a reduction factor $R_{heat}$. This provides a metric to assess the quality of a robust PST. For all schemes discussed in the following sections, decreasing $R_{heat}$ comes at the cost of decreasing the gate speed or increasing the sideband powers. In order to provide a useful comparison, the efficiency of schemes are compared for fixed powers. It is therefore useful to introduce a gate time scaling cost $R_{time} = \tau / \tau_0$. An efficient scheme is one that provides a low reduction factor $R_{heat}$ while keeping the time cost $R_{time}$ small. We introduce a final scaled heating reduction factor, $\tilde{R}_{heat} = R_{heat} R_{time}$, such that the infidelity of equation \ref{eq:reduction_factor_heating_simple} becomes

\begin{equation}\label{eq:infid_heating_scaled_reduction_factor}
\mathcal{I} = \frac{1}{2}\ndot \tilde{R}_{heat} \tau_0.
\end{equation}

This final factor $\tilde{R}_{heat}$ more accurately represents the performance and will prove useful for benchmarking schemes. 

We have formulated robustness to motional decoherence as a geometric problem in which one must engineer PSTs that satisfy the following requirements: (i) the PST must end at the origin at the completion of the gate ($\mathcal{R}_{end}$), (ii) the average displacement should be zero ($\mathcal{R}_{dephasing}$) and (iii) one should minimize the average distance to the origin ($\mathcal{R}_{heating}$). Note that if the PST is symmetric about an arbitrary axis, the requirement of (i) is automatically satisfied by (ii). Symmetrization can therefore alleviate constraints during numerical optimisations, as well as reduce the total number of parameters. In the following sections, we discuss various methods that can engineer an arbitrary PST.

%% file: _sections/Section2/pst_engineering.tex
\subsection{Engineering a phase space trajectory} \label{sec:engineering_arbitrary_pst}

\subsubsection{Sideband Modulation} \label{sec:sideband_modulation}

\begin{figure}[t]
\center
\includegraphics[scale=.97]{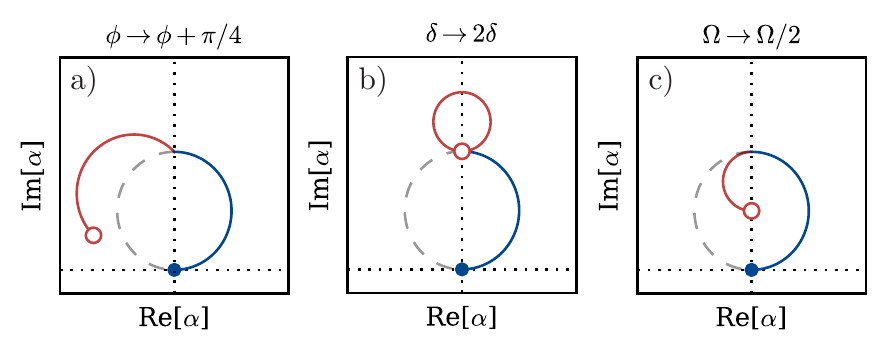} 
\caption[The effects of phase, amplitude and frequency modulation on phase space trajectories.]{Engineering phase space trajectories with phase, frequency and amplitude modulation (PM, FM and AM). The PST starts at the origin (full circle) and ends at the white circle. During the first half of the evolution (blue), the sideband parameters are kept constant. The phase (a), frequency (b) or amplitude (c) are changed instantaneously and kept constant throughout the second half (red). (a) A phase change instantaneoulsy changes the direction of the PST, however the angular velocity and the radius of the arc are preserved. (b) A change in frequency not only changes the direction, but also the arc radius. (c) Amplitude modulation affects both the angular velocity and the arc radius. The path length is therefore smaller (or longer). }
\label{fig:modulation_comparison}
\end{figure}

An arbitrary PST can be implemented by modulating one or more of the parameters that make up the path $\alpha_k(t)$ of equation \ref{eq:generalized_pst}. Any combinations of amplitude, phase and frequency modulation can be used, either through continuous or piecewise constant functions \cite{wright2019, blumel2019,murali2020, ball2021}. The effects of an instantaneous change in the parameters are illustrated in figure \ref{fig:modulation_comparison}. Phase modulation (PM) rotates the displacement vector, and the centres of rotation of the subsequent PST are displaced. Frequency modulation (FM) affects the curvature of the PST. Finally, amplitude modulation (AM) varies both the curvature of the PST and the strength of the interaction. The angular velocity of the PST is therefore affected by changes in the amplitude.

Robust gates with AM were first demonstrated in \cite{choi2014} with a five ion chain using laser beams. A microwave based AM gate was later demonstrated in \cite{zarantonello2019}, where the amplitude was assumed to have a sinusoidal envelope of order $n$ such that $\Omega(t) = \Omega_0\sin^n(\alpha t)$. The duration of a robust gate of order $n=2$ is twice as long. AM gates have since been used to achieve Efficient, Arbitrary, Simultaneously Entangling (EASE) gates in large trapped ion registers \cite{grzesiak2020}.

FM gates were first demonstrated with laser beams \cite{leung2018}. The frequency is modelled as a continuous sinusoidal oscillation with optimizable vertices. A similar sequence was adopted in \cite{leung20182}, where an additional slow sinusoidal ramp was applied to the amplitude. Alternatively, a discrete sequence of equidistant frequencies can be used for compatibility with Direct Digital Synthesis boards \cite{wang2020}. Finally, FM was used to demonstrate parallel entangling gates with arbitrary pairs of ions in a long chain \cite{landsman2019}, and the PSTs can be made even more robust via batched numerical optimizations \cite{kang2021}. 

Following the initial proposal of \myref \cite{green2015}, PM gates have been demonstrated with laser beams \cite{milne2020}. A sequence of up to 32 discrete phases demonstrated robustness to both static and time-varying noise. PM was also used to achieve global entangling gates in larger ion chains \cite{lu2019}.

Robust modulated gates have been widely adopted across a variety of trapped-ion architectures. The choice, however, between employing AM, PM or FM gates largely depends on the application. Furthermore, microwave and laser gates may have different preferences. A first criterion is the scaling of the gate time. For larger ion chains ($N\gg 2$), the total time of AM and PM gates scales linearly with the distance between the ion pair since non-neighbour interactions are weaker \cite{murali2020}. The duration of FM gates, however, scales linearly with the length of the ion chain. The different scalings lead to vastly different performances for various algorithms within large chains and it is therefore crucial to make the appropriate selection \cite{murali2020}. Nevertheless, in what follows we only consider small ion chains that would be used within a QCCD type architecture, such as the one considered in Ref. \cite{lekitsch2017}.

\input{_sections/Section2/mtms_table.tex}

A variety of modulated sideband schemes were also developed with the aim of mitigating errors from off-resonant coupling to other motional states. In laser-based architectures with large strings of ions, the motional frequencies are spectrally crowded and spectator modes strongly interact with the gate field. The modulation sequences are therefore designed such that the total entangling phase from every contributing motional mode leads to maximal entanglement \cite{milne2020, qctrl_molmer_sorensen}. Furthermore, the sequence ensures that all modes of vibration are decoupled from the spin states at the completion of the gate. These restrictions are fortunately alleviated for a QCCD architecture with global microwave fields and static magnetic field gradients. On the one hand, the small Lamb-Dicke parameter strongly suppresses coupling to spectator modes of vibration. On the other hand, the achievable Rabi frequencies with global radiation fields are smaller than for laser beams. The resulting infidelity is on the order of $10^{-4}$ to $10^{-6}$ (see appendix \ref{app:off_res_coupling_motion}). For this reason, the sideband modulation schemes will only be discussed in the context of improving robustness towards motional decoherence. The additional constraint of managing parasitic couplings to spectator modes is not an issue for the architecture considered here, which grants more freedom in designing efficient PSTs.

\subsubsection{Multi-Tone M{\o}lmer-S{\o}rensen Gate} \label{sec:mtms}

In the previous section, it was shown that arbitrary PSTs can be engineered by modulating the parameters of the sideband fields. Alternatively, one can keep the sideband fields constant and add bichromatic tones of varying amplitudes, which also contribute to the \msgate\ interaction \cite{haddadfarshi2016, webb2018, shapira2018}. In this way, an $N$ tone gate involves $2N$ sideband fields per ion. The detuning of the $j^{th}$ bichromatic fields are $\delta_j = j \delta_0$ and their amplitudes are $\Omega_j = c_j \Omega_0$. The real valued coefficients $c_j$ greatly influence the dynamics of the interaction and are chosen to implement robust sequences. The Hamiltonian describing an $N$-tone \msgate\ gate is (setting $\phi = 0$)

\begin{equation}
H = \frac{\hbar\epsilon\Omega_0}{2}S_x \sum_{j=1}^N c_j(a\dag e^{ij\delta_0 t} + a e^{-ij\delta_0 t}),
\end{equation}
with $S_x = \sigma_x^{(1)} + \sigma_x^{(2)}$, and the PST is

\begin{equation}
\alpha_k(t) = \sum^N_{j=1} \alpha_{j,k}(t),
\end{equation}
where $\alpha_{j,k}(t)$ is the displacement of the $j^{th}$ tone with Rabi frequency $\Omega_j$ and detuning $\delta_j$. By choosing appropriate coefficients $c_j$, an $N$-tone MS gate can implement robust PSTs. The coefficients are first constrained by setting $\sum c_j^2/j = 1$, which is required to generate a maximally entangled state. The zero-averaged position requirement of equation \ref{eq:pst_requirement_average_center} ($\mathcal{R}_{dephasing}$) is further satisfied by setting $\sum c_j/j = 0$. Finally, the quantity $\sum c_j^2/j^2$ should be minimized as per the requirement of equation \ref{eq:pst_requirement_min_heating} ($\mathcal{R}_{heating}$) which leads to gates that are robust to heating. The reduction factor of the heating rate that was defined in equation \ref{eq:reduction_factor_heating} now becomes 

\begin{equation} \label{eq:mtms_r_heating}
R_{heat} = \frac{\ndot_{eff}}{\ndot} = \frac{1}{2}\sum_j \frac{c_j^2}{j^2}.
\end{equation}

An optimal set of coefficients are derived from the previously mentioned constraints, resulting in $c_j = 4\frac{jb}{1 - j\lambda}$ where $b = -\frac{1}{4}(\sum_j \frac{j}{(1-j\lambda)^2})^{-1/2}$ and $\sum_j 1/(1-j\lambda) = 0$ \cite{haddadfarshi2016}.

It is also interesting to note the power requirements of an $N$-tone gate. Given the close spectral proximity of the tones ($\delta_0$) and the time-scale of the interaction ($2\pi/\delta_0$), additive and destructive interferences introduce a time-dependence on the amplitude. The total signal is modelled as $f_N(t) = \sum^N_j c_j e^{i (\omega + j\delta_0)t}$, and the maximum amplitude of an $N$-tone gate is $\textrm{max}(|f_N(t)|)_{t\in [0, \tau]}$ \cite{lishman2020}. Constraints on the maximum amplitude of the physical fields arise from classical hardware limitations. Therefore, the primitive Rabi frequency $\Omega_0$ of a multi-tone gate should be adjusted, as both the amplitude and gate duration are scaled by $\textrm{max}(|f_N(t)|)$. For example, a 2-tone gate involves two red (blue) sideband fields with frequency separation $\delta_0$. The amplitude of the total signal is $|f_N(t)| = \sqrt{(5 - 4\cos(t\delta_0))/3}$, which is maximized at half the gate duration where $|f_N(\tau/2)| = \sqrt{3}$. The peak amplitude is therefore $\sqrt{3}$ times larger compared to a primitive \msgate\ gate and the gate duration and Rabi frequency should be adjusted by $\tau \rightarrow \sqrt{3}\tau $ and $\Omega_0 \rightarrow \Omega_0 /\sqrt{3}$ to satisfy the power requirements. The gate time scaling of an $N$-tone gate is

\begin{equation} \label{eq:mtms_r_time}
R_{time} = \textrm{max}(|f_N(t)|)_{t\in [0, \tau]}.
\end{equation}

The coefficients, heating and gate time scalings for up to $N=5$ tones are reported in table \ref{tab:mtms_coeffs}. For increasing numbers of tones, $R_{heat}$ decreases and leads to larger reductions of the heating rate. For example, a 5-tone gate leads to a tenfold improvement. However, the time scalings also progressively worsen, and the tenfold improvement comes at the cost of a gate that is $2.66\times$ slower. The time-scaled heating reduction factors $\tilde{R}_{heat}$ are also reported. After considering the longer gate duration, the tenfold improvement of a 5-tone gate reduces to a threefold reduction of the heating rate.

The multi-tone interaction has also been used for various other interesting applications, which aren't considered here. For example, it was shown that adding multiple tones allows one to operate the \msgate\ interaction in the nonadiabatic regime, effectively coupling all motional modes and achieving high-fidelity gates within large ion crystals \cite{shapira2020}. Alternatively, the amplitude coefficients can be numerically optimized to increase the robustness of the interaction to qubit frequency errors \cite{lishman2020}.

%% file: _sections/Section2/mtms_table.tex
\begin{table*}[th!]
\center
\begin{tabular}{ccccccccc}
\hline\hline
 Tones & $c_1$ & $c_2$ & $c_3$  & $c_4$& $c_5$ &  $R_{heat}$ & $R_{time}$ & $\tilde{R}_{heat}$\\ \hline
1 &  1  & 0   &  0 & 0 & 0 & 1  & 1 & 1 \\
2 &  $-1\sqrt{3}$  & $2/\sqrt{3}$  & 0 &  0  & 0& 0.33  &   1.73 &  0.57\\
3 &  -0.132  &  -0.719  & 1.474  & 0 & 0 & 0.19  &  2.06 & 0.40\\ 
4 & -0.06 & -0.204 & -0.804 & 1.74 & 0 & 0.14 & 2.4 & 0.33\\
5 & -0.036 & -0.104 & -0.256 & -0.872 & 1.972 & 0.11 & 2.66 & 0.28\\
\hline\hline
\end{tabular}
\caption[Optimal coefficients and performance of an $N$-tone \msgate\ gate.]{Optimal coefficients and performance of an $N$-tone \msgate\ gate. The factors $R_{heat}$ and $R_{time}$ are calculated from \protect\eqref{eq:mtms_r_heating} and \protect\eqref{eq:mtms_r_time} respectively. The time-scaled heating reduction factor is $\tilde{R}_{heat}= R_{heat} R_{time}$.}
\label{tab:mtms_coeffs}
\end{table*}

%% file: _sections/Section2/comparison.tex
\subsubsection{Comparison}

It was shown in section \ref{sec:robust_psts} that robustness to motional decoherence is entirely dependent on the PST, regardless of the physical implementation. Therefore, if two schemes engineer the same trajectory, the expected robustness is identical. An important difference, however, is the gate time scalings, the physical resources of a scheme and their experimental complexity. In what follows, these factors are used to compare the sideband modulation schemes (section \ref{sec:sideband_modulation}) with the multi-tone \msgate\ gate (section \ref{sec:mtms}).

\begin{figure}[t]
\center
\includegraphics[scale=1]{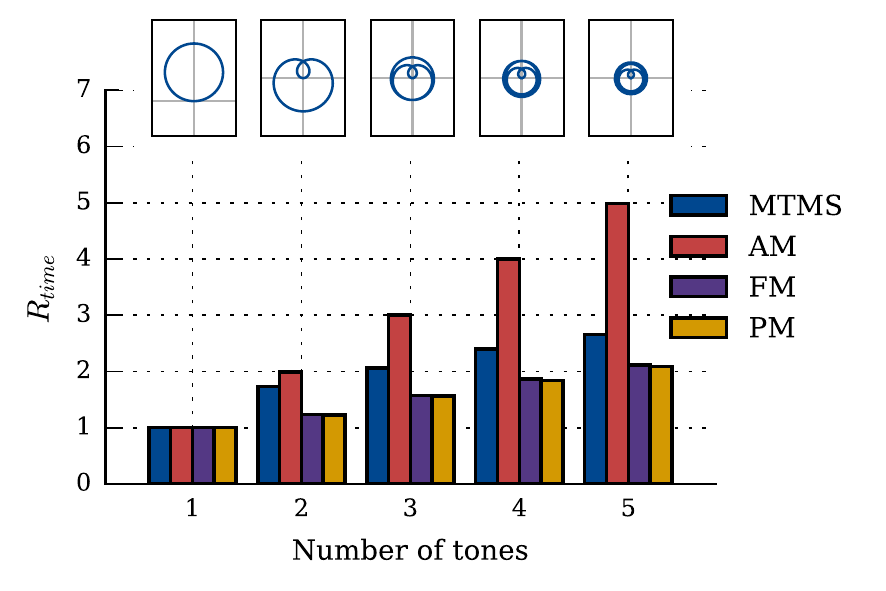} 
\caption[Comparison of gate time scalings for a PM, FM, AM and a multi-tone \msgate\ gate.]{Comparison of gate time scalings $R_{time}$ for a phase modulated (PM), frequency modulated (FM), amplitude modulated (AM) and multi-tone (MTMS) \msgate\ gate. The modulation sequences are chosen such that the resulting PST implements an $N$-tone gate. The corresponding PSTs are plotted in the insets. }
\label{fig:motional_robust_gates_comparison_time_scaling}
\end{figure}

While sideband modulation can engineer an arbitrary path in phase space, the PST of multi-tone gates are fixed by the optimized coefficients. The schemes are therefore compared with one another by finding sideband modulation sequences which implement an $N$-tone \msgate\ gate. In this way, the gate time scalings of the modulation schemes can be compared to those reported in table \ref{tab:mtms_coeffs}. The modulation sequences are found from numerical optimisations, which are detailed in appendix \ref{app:mod_seq_mtms}. The gate time scalings for each scheme are reported in figure \ref{fig:motional_robust_gates_comparison_time_scaling}. We first notice that AM is the least efficient  as it results in the largest time scalings. This is expected given that an AM sequence implements a PST by modulating the gate speed with an upper bound being $\Omega_{max} = \Omega_0$. The FM and PM sequences have near identical scalings given that they are equivalent ($\delta(t) = \dot{\phi}(t)$). Since the amplitude is kept at $\Omega_0$, the gate time cost is purely due to the smaller area enclosed by the PST. Furthermore, the FM and PM sequences perform better than an MTMS gate for any number of tones. For example, the FM and PM sequences are $1.4\times$ faster than a 2-tone gate, with $R_{time}^{FM} = R_{time}^{PM} = 1.23$ and $R_{time}^{MTMS} = 1.73$.

We now discuss the physical requirements and experimental complexity of the schemes. An MTMS gate requires $4N$ additional fields (two sideband fields per ion for each tone due to the static magnetic field gradient). The experimental complexity quickly increases, as each new tone requires a precise calibration of its amplitude. Conversely, the FM and PM sequences do not introduce any additional fields, and are implemented programatically with arbitrary waveform generators. Due to the high phase resolution, additional calibrations are in principal not required. The FM and PM sequences differ when considering the functional forms of their modulation. Robust PSTs engineered with PM use piecewise-constant functions, while FM uses continuously varying functions. 

We conclude that AM and PM are more efficient as they lead to the smallest gate time scalings. Furthermore, their experimental complexity is minimal compared to multi-tone gates. In the following sections, only PM is considered as it is conceptually simpler to work with piecewise constant functions.

%% file: _sections/Section2/optimal_pst.tex
\subsection{Optimal phase space trajectories} \label{sec:optimal_pm_pst}

Robust PSTs suffer from a smaller enclosed area, hence a prolonged gate duration is required for maximal entanglement. It is interesting to explore this trade-off and characterize the performances of optimal PSTs, \ie trajectories which maximise the enclosed area for a given target robustness, thereby minimising the gate time scaling. We first consider the robustness to motional heating and build a model which, provided a desired reduction factor $R_{heat}$, estimates the smallest gate time scaling $R_{time}$ that is achievable from an optimised PM sequence. 

\newcommand{\avdist}{\langle|\alpha(t)|^2\rangle}

\begin{figure}[t]
\center
\includegraphics[scale=1]{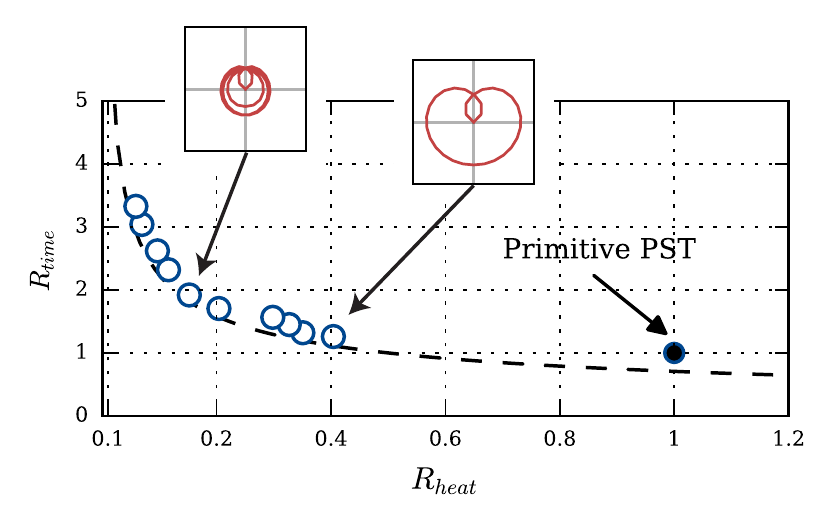} 
\caption[Trade-off between robustness to motional heating and gate duration.]{Trade-off between robustness to motional heating and the gate duration. Each data point corresponds to a PST obtained by a numerical optimisation algorithm. The enclosed area of the PSTs are maximised, such that the gate time scaling $R_{time}$ is minimised. The insets illustrate two examples of optimised PSTs. The performance of a primitive \msgate\ gate is included for comparison. }
\label{fig:time_vs_distance}
\end{figure}

A library of PM sequences is created such that each resulting PST achieves a certain reduction factor $R_{heat}$. Each PM sequence is the result of a numerical optimization which minimizes the scaling $R_{time}$. In this way, the resulting PST for a given $R_{heat}$ corresponds to the fastest solution. In practice, the numerical optimisation algorithm maximises the enclosed area of the PST while constraining its average distance from the centre. The optimisations are further constrained by ensuring symmetry, closure and a zero-averaged position of the PSTs (see the requirements detailed in \ref{sec:robust_psts}). The time scalings $R_{time}$ are calculated by numerically simulating the \msgate\ Hamiltonian and minimising the resulting infidelity. The heating reduction factors $R_{heat}$ are then computed from equation \ref{eq:reduction_factor_heating}. The heating reduction factors and time scalings of the resulting PM sequences are reported in figure \ref{fig:time_vs_distance}. Interestingly, the fastest PST which satisfies the aforementioned constraints achieves a heating reduction of $R_{heat} = 0.4$ with a time scaling $R_{time} = 1.2$ ($\tilde{R}_{heat} = 0.48$). Conversely, the most robust PST that was simulated with $R_{heat} = 0.06$ results in a time scaling $R_{time} = 3.3$ ($\tilde{R}_{heat} = 0.2$). 

In order to model the trade-off between robustness and time cost, the optimised PSTs are approximated by $m$ circles centred at the origin (see inset of figure \ref{fig:time_vs_distance}). The average distance is therefore constant and equal to the radius of the circles $r' = (\avdist)^{1/2}$. In order to accumulate a maximally entangling phase, the area $m\pi (r')^2$ enclosed by $m$ circles must be equal to that of a primitive \msgate\ gate, \ie $\mathcal{A}_0 = \pi r_0^2 $ with $r_0 = 2\sqrt{2}$. For an identical gate time, a PST with a smaller average distance will enclose a smaller area, however the length of its trajectory is conserved. Therefore, the perimeter of the $m$ circles must be equal to that of the primitive PST, \ie $\mathcal{P}_0 = 2\pi r_0 = m 2\pi r''$. With these previously defined relations, the theoretical gate time scaling is found from $R_{time} = \tau/\tau_0 = \sqrt{\mathcal{A}_0/\mathcal{A}}$ where $\mathcal{A} = \pi r''^2$,

\begin{equation} \label{eq:av_dist_to_rtime}
R_{time} = (2 R_{heat})^{-1/2}.
\end{equation}

The theoretical time scalings are plotted alongside the numerically optimised PSTs in figure \ref{fig:time_vs_distance} and show good agreement. Deviations are attributed to the finite path length that is required to join the PST to the origin. The model of equation \ref{eq:av_dist_to_rtime} is now combined with equation \ref{eq:infid_heating_scaled_reduction_factor} to find the infidelity for an optimal PST,

\begin{equation} \label{eq:infid_model_rheat}
\infid = \frac{1}{2} \ndot \sqrt{\frac{R_{heat}}{2}} \tau_0.
\end{equation}

\begin{figure}[t]
\center
\includegraphics[scale=1]{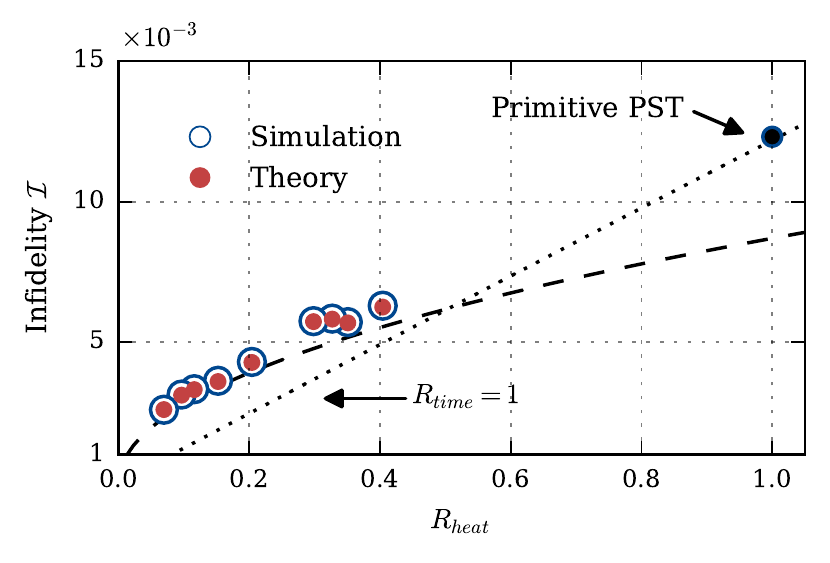} 
\caption[Infidelity of the \msgate\ entangling gate to motional heating for various robust phase space trajectories.]{Infidelity of the \msgate\ gate to motional heating for various robust PSTs. Numerical simulations considered a heating rate $\ndot = \SI{40}{s^{-1}}$, a Lamb-Dicke parameter $\epsilon = 0.01$, and an MS Rabi frequency $\Omega/2\pi = \SI{40}{kHz}$, resulting in a gate duration of $\tau = \SI{1.25}{ms}$ for a primitive PST (single loop). For robust schemes, the gate duration in numerical simulations is scaled by $R_{time}$. Theory data points are calculated from \protect\ref{eq:infidelity_heating} using the PST's computed $R_{heat}$ and $R_{time}$. The dotted line corresponds to this same infidelity term after setting $R_{time} = 1$, and therefore represents the infidelity if the gate duration were kept constant. The dashed line is obtained from the theoretical model of \protect\ref{eq:infid_model_rheat}. Finally, the primitive PST corresponding to a simple single-loop \msgate\ gate is plotted for reference.}
\label{fig:heating_vs_time_tradeoff}
\end{figure}

This infidelity model is validated by numerically integrating the \msgate\ Hamiltonian with the optimized PM sequences. Heating is integrated by incorporating the usual Linbladian operators in the Master Equation \cite{sorensen2000}. For each PST, the gate time and detuning are adjusted with the transformations $\tau \rightarrow R_{time}\tau$ and $\delta \rightarrow \delta/R_{time}$. The results are reported in figure \ref{fig:heating_vs_time_tradeoff}. The analytically predicted infidelities are calculated from \ref{eq:infid_heating_scaled_reduction_factor} by using the exact $R_{heat}$ and $R_{time}$. Both simulation and predictions match well, which validates the approach of modelling infidelities from purely geometric properties of the PST. Furthermore, the idealized model of \ref{eq:infid_model_rheat} is plotted alongside the results and shows good agreement. The infidelity model for robustness to heating can be incorporated into the spin robustness model derived in \ref{sec:robustness_spin_dephasing} after replacing $\tau \rightarrow R_{time} \tau$.

Having investigated the efficacy of engineered PSTs at mitigating heating errors, we now move our attention towards imbalances in the \msgate\ detuning. By design, the PSTs are to first order insensitive to errors in the detuning (c.f. $\mathcal{R}_{dephasing}$ of equation \ref{eq:pst_requirement_average_center}). Therefore, the quality of a PST at mitigating detuning errors is assessed from its quadratic sensitivity,

\begin{equation} \label{eq:quadratic_sensitivity_pst}
\frac{\partial^2 \alpha (\tau)}{\partial \delta^2}\bigg|_{\delta=0} = -\tau^2 \alpha(\tau) + 2\tau \alpha_{av}(\tau) - 2 \int^{\tau}_0 dt \alpha_{av}(t),
\end{equation}

where we recall that the average displacement in phase space is $\alpha_{av}(t) = \int^{t}_0 dt' \alpha(t')$. The first two terms of equation \ref{eq:quadratic_sensitivity_pst} reduce to zero for a robust sequence that satisfies the requirements of equations \ref{eq:pst_requirement_finish_center} and \ref{eq:pst_requirement_average_center}. The last term, representing the time-averaged mean position in phase space, is then used as a metric of comparison.

To verify the robustness of PSTs to detuning errors, equation \ref{eq:quadratic_sensitivity_pst} is evaluated for the two PSTs that are shown in the insets of figure \ref{fig:time_vs_distance}. The PST that is more robust to heating is found to have a quadratic sensivitiy that is $3.7 \times $ smaller. In general, one can make the assumption that robustness to heating and detuning errors go hand in hand, since minimizing the distance to the origin also reduces the time-averaged mean position.

%% file: _sections/Section2/thermal_noise.tex
\subsection{Robustness to thermal noise}

The \msgate\ interaction is in principle insensitive to the initial temperature of the modes of motion. From equation \ref{eq:off_res_coupling_spectator_mode}, one can see that the infidelity is zero provided that the PST returns to the origin. However, higher motional temperatures amplify errors due to non-closures of the PST. Therefore, in realistic settings where small parameter missets are present, higher temperatures worsen the effects of experimental imperfections. Aside from errors in the PST, thermal noise also amplifies infidelities from qubit frequency missets such as those considered in \ref{sec:robustness_spin_dephasing}. 

\begin{figure}[t]
\center
\includegraphics[scale=.9]{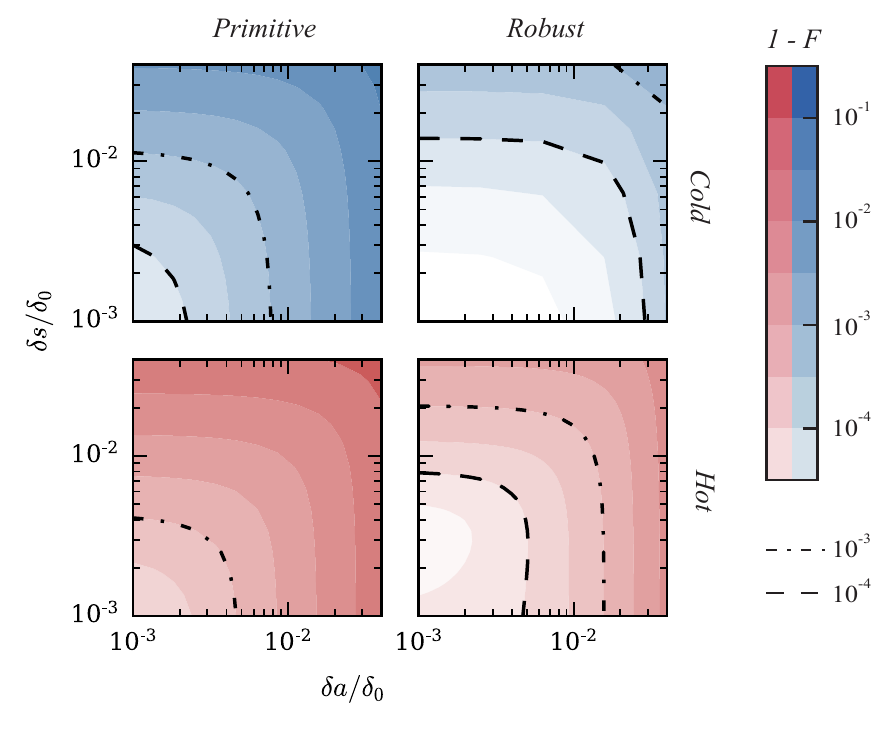} 
\caption{Numerically simulated infidelities of a primitive and robust (CDD and PM) \msgate\ entangling gate for cold (blue, $\nbar = 0$) and hot (red, $\nbar = 5$) initial temperatures. Their robustness is characterized by introducing static frequency missets in both the qubit frequency ($\delta_a$) and motional mode frequency ($\delta_s$). The dashed (dash-dotted) lines represent infidelities of $10^{-4}$ ($10^{-3}$). }
\label{fig:infidelity_vs_temperature}
\end{figure}

The extensions of the bichromatic interaction (such as CDD for spin robustness and PM for motional robustness) can be used to obtain high fidelities in the presence of large temperatures. These "hot" gates may alleviate challenging experimental requirements such as sideband cooling. The increased robustness in the presence of thermal noise is reported in figure \ref{fig:infidelity_vs_temperature}. Both a primitive and robust entangling gate are numerically simulated for hot ($\nbar=5$) and cold ($\nbar=0$) temperatures. Here, the robust scheme consists of CDD with PM on the sideband fields. In this way, the resulting interaction is resilient to both spin and motional decoherence. One can see that a hot robust gate achieves higher fidelities than a cold primitive gate for similar parameter missets. This increased robustness is expected to hold for higher temperatures, such as ones found in ion strings that are only cooled to their Doppler limit.

%% file: _sections/Section2/summary_and_outlook.tex
\subsection{Outlook}

The previous sections have identified extensions to the \msgate\ gate which provide robustness to both spin and motional decoherence. While these can be used to independently increase the coherence of either internal or external degrees of freedom, they can more importantly be combined to provide simultaneous robustness. For example, phase modulation of the sideband fields can be combined with a continuous dynamical decoupling carrier field. A proof of concept demonstration of this particular interaction is presented in section \ref{sec:experimental_demonstration}. Another example of such combinations can be found in Ref. \cite{arrazola2020}, wherein the \msgate\ interaction is subject to continuous dynamical decoupling and additional $\pi$-pulses that lead to motional robustness. 

More generally, the various quantum control schemes previously outlined make up a library of gate extensions that can be combined with one another. The particular choice of which schemes to use is then motivated by the desired performance of an experimental system and its available hardware. For example, if the fidelity of the entangling operation should be maximized, one can build an error model from the infidelity terms derived in the previous sections and optimize over parameters such as the gate duration or the dynamical decoupling Rabi frequency. Furthermore, the gate scheme library may be constrained by factors such as the available polarisation of the microwave fields or their achievable bandwidth. 

%% file: _sections/Section3/section3.tex
\section{Experimental demonstration} \label{sec:experimental_demonstration}

\input{_sections/Section3/introduction.tex}

\input{_sections/Section3/robustness_spin.tex}
\input{_sections/Section3/robustness_motion.tex}

%% file: _sections/Section3/introduction.tex
\begin{figure}[t]
\center
\includegraphics[scale=.9]{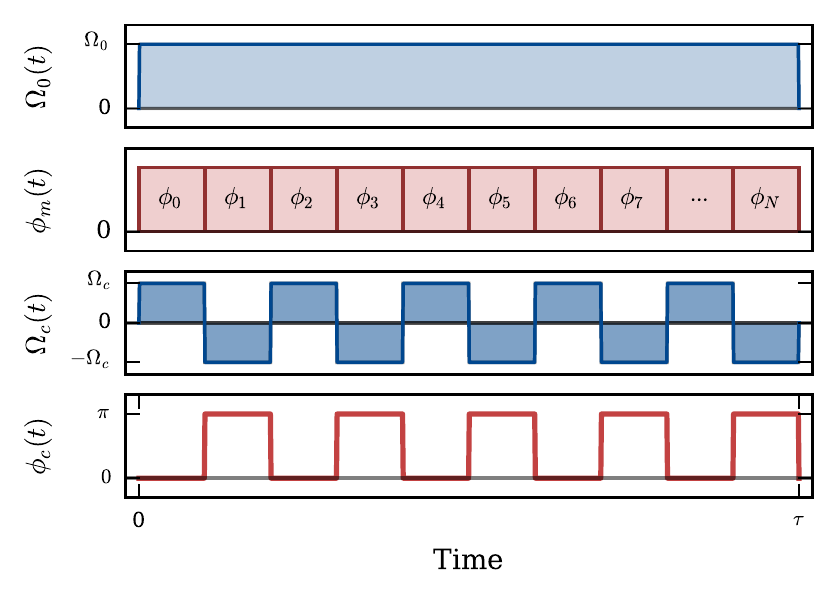} 
\caption{Pulse sequence for a \msgate\ gate that is simultaneously robust to spin and motional decoherence. Both the sideband fields ($\Omega_0$, $\phi_m$) and the carrier ($\Omega_c$, $\phi_c$) are represented. Phases are represented in red and amplitudes in blue. The motional phase is modulated by a piecewise constant function. The sideband Rabi frequency is kept constant throughout the gate interaction. The carrier's phase alternates between 0 and $\pi$, causing a change in the sign of the carrier Rabi frequency. The frequency with which phase flips occur are determined from the number of motional phase modulation segments.}
\label{fig:gate_sequence}
\end{figure}

Sections \ref{sec:robustness_spin_dephasing} and \ref{sec:robustness_motional_decoherence} outlined a number of quantum control schemes that extend the robustness of the primitive \msgate\ interaction to either spin or motional decoherence. In principle, multiple schemes can be combined to enable a simultaneous robustness to both sources of error. For example, one could extend the spin's coherence via PDD, while increasing the robustness to motional heating through frequency modulation of the sideband fields (although extra care should be taken in keeping track of additional phase offsets introducing by the $\pi$-pulses). 

Here, we experimentally demonstrate the construction of a robust gate from a library of quantum control schemes by combining \textit{continuous dynamical decoupling}, \textit{rotary echoes} and \textit{sideband phase modulation}. The single-ion equivalent of the bichromatic \msgate\ interaction is used as a proof-of-concept experiment. The qubit is encoded within the $\{\ket{F=0,m_f=0}, \ket{F=1, m_f=1}\}$ states of the $^2S_{1/2}$ hyperfine ground state of a $^{171}$Yb$^+$ ion. All fields involved in the interaction are in the microwave regime, with a frequency nearing $\SI{12.64}{GHz}$ (see appendix \ref{app:experimental_details} for further experimental details). The pulse sequence is illustrated in figure \ref{fig:gate_sequence}. The sideband fields are applied with a square pulse at a fixed Rabi frequency $\Omega_0$. The phases of the red and blue sidebands, however, are subject to a discrete phase modulation sequence. The amplitude of the dynamical decoupling carrier field is constant, however phase flips are introduced after every sideband phase change to implement rotary echoes.

%% file: _sections/Section3/robustness_spin.tex
\subsection{Robustness to spin decoherence}

\begin{figure}[t]
\center
\includegraphics[scale=.9]{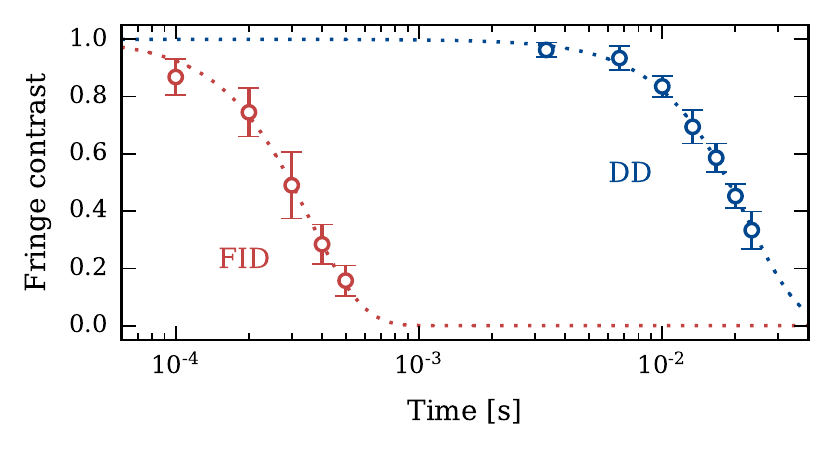} 
\caption{Coherence time measurement of a free induction decay experiment (red) and continuous dynamical decoupling subject to rotary echoes (blue). Dotted lines are fits to a Gaussian decay.}
\label{fig:experimental_spin_robustness}
\end{figure}

The increased robustness to spin decoherence is first characterized. In the absence of any dynamical decoupling, the coherence time is measured to be $T_2^*=\SI{358(8)}{\mu s}$ via a Free Induction Decay (FID) Ramsey experiment (see figure \ref{fig:experimental_spin_robustness}). The coherence time is dominated by magnetic field noise which couples to the magnetically sensitive transition. We then add a continuous carrier field with a Rabi frequency $\Omega_c/2\pi = \SI{29.9}{kHz}$, and find that the driven coherence time is on the order of $\SI{500}{\mu s}$. It was found that decoherence during driven evolution was dominated by amplitude noise in the carrier field itself. This is further mitigated by introducing rotary echoes, \ie phase flips which alternate the carrier Rabi frequency from $\Omega_c$ to $-\Omega_c$ and refocus amplitude noise. The phase flips are applied after every $2\pi$ rotation of the carrier, i.e. with a period of $\SI{33.44}{\mu s}$. The coherence time under CDD with rotary echoes is finally measured via a Ramsey-type experiment, resulting in $T_2 = \SI{22.6(3)}{ms}$. Continuous dynamical decoupling therefore extends the spin's coherence by almost two orders of magnitude, despite significant noise in the carrier field itself.

%% file: _sections/Section3/robustness_motion.tex
\subsection{Robustness to motional decoherence}

The robustness to motional decoherence is verified by creating a Schr\"odinger cat state. After initialisation of the qubit, the bichromatic fields are applied and result in a state dependent motional displacement, whose trajectory in phase space is identical to that of a multi-qubit entangling gate. For an initial state $\ket{\downarrow} \otimes \sum_n p_{\nbar} (n)\ket{n}$, where $p_{\nbar}(n)$ is the Maxwell-Boltzmann distribution, the probability of measuring the $\ket{\uparrow}$ state is 

\begin{equation} \label{eq:schrodinger_cat_probability}
P_{\uparrow} = \frac{1}{2}(1 - e^{-2 |\alpha(t)|^2(1+2\nbar)}).
\end{equation}

Under the correct \msgate\ detuning, the phase space trajectory $\alpha(t)$ returns to zero at the gate duration $\tau$, and therefore $P_{\uparrow} = 0$. Errors such as parameter missets or motional decoherence directly result in a non-zero spin probability. Therefore, the robustness of the PST can be inferred from measurements of $P_{\uparrow}$.

Robustness to motional decoherence is demonstrated by comparing the probabilities resulting from a primitive bichromatic interaction to one with sideband PM. The sideband fields are set to $\Omega_0/2\pi = \SI{30}{kHz}$. The bichromatic detuning of the primitive \msgate\ interaction is $\delta_0/2\pi = \SI{321}{Hz}$, resulting in the gate duration $\tau = \SI{3.12}{ms}$. Note that at time $\tau$, the interaction picture of the bichromatic interaction and the carrier fields must coincide. In other words, the carrier dynamical decoupling field should have completed an integer number of $2\pi$ rotations. This is satisfied by setting $\Omega_c = N \delta_0$, and the carrier Rabi rate was chosen such that $N=93$. This further constrains the efficiency of the rotary echoes, as one can now only implement $N=93$ phase flips on the carrier.

A robust PST is obtained via numerical optimizations \cite{ball2021, qctrl_molmer_sorensen} and obeys all robustness conditions outlined in section \ref{sec:robust_psts}. The resulting gate time scaling is $R_{time} = 1.218$. The modified bichromatic detuning is $\delta'/2\pi = \SI{264}{Hz}$, with the gate duration $\tau' = \SI{3.79}{ms}$. This corresponds to a phase sequence of $N=114$ segments.

\begin{figure}[t]
\center
\includegraphics[scale=1]{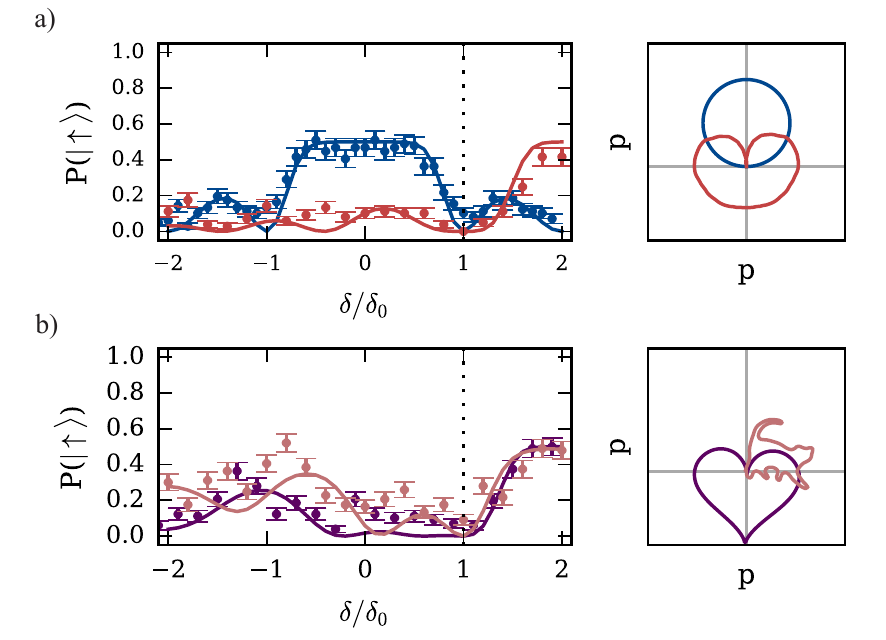} 
\caption{Measurement probability of the ion being in the $\ket{\uparrow}$ state after the creation of Schr\"odinger cat states for varying bichromatic detunings. The correct phase space trajectory is obtained for detunings intersecting with the dotted line. Solid lines are analytical predictions calculated from equation \ref{eq:schrodinger_cat_probability}. (a) A primitive (blue) and robust (red) sequence are considered. The PST of the latter is constructed to obey all robustness requirements. (b) Phase modulation sequences whose PSTs implement a cat (pink) and a heart (purple). We wish to dedicate this heart-shaped PST and the resulting measured state probability to Bruce W. Shore whose kindness and scientific genius enriched the field of coherent control and scientists working in the field in numerous ways.}
\label{fig:experimental_motion_robustness}
\end{figure}

The primitive and robust interactions are compared in figure \ref{fig:experimental_motion_robustness}. The fields are applied for a duration $\tau$ ($\tau'$) for the primitive (robust) sequence and the bichromatic detuning is varied from $-2\delta_0$ to $2\delta_0$ ($-2\delta_0'$ to $2\delta_0'$). The correct interaction of the primitive (robust) sequence is obtained for $\delta = \pm \delta_0$ ($\delta = \delta_0'$). Around these detunings, the response of the measured probability differs between both sequences. The primitive interaction exhibits a narrow region, while the robust interaction results in a flatter region. The latter is therefore more tolerant to missets in the bichromatic detuning, and by extension is more robust to motional decoherence mechanisms. This is further verified by noting that its PST satisfies all the requirements for robustness (\cf figure \ref{fig:experimental_motion_robustness}): (i) the path is symmetric about an axis, (ii) the average position is zero and (iii) the average distance to the centre is smaller. Finally, it is interesting to note that the experimentally measured probability $P_\uparrow$ of the robust sequence is smaller than for the primitive sequence, as this is a reflection of the interaction's increased resilience.

An important advantage in using microwave radiation for scalable quantum computing is the resolution and stability of off-the-shelf components. Commercial microwave sources can achieve very low amplitude and phase noise. Furthermore, non-linearities in the synthesis chain are dominated by components such as amplifiers, whose effects can be made small. To demonstrate this ease of control, we experimentally create a Schr\"odinger cat state by drawing out a cat trajectory in phase space (see figure \ref{fig:experimental_motion_robustness}). Furthermore, considering the work of Ref. \cite{shapira2018}, we take the literal sense of the  PSTs dubbed "Cardiods" and demonstrate a heart-shaped trajectory. 

We wish to dedicate this heart-shaped PST and the resulting measured state probability to Bruce W. Shore whose kindness and scientific genius enriched the field of coherent control and scientists working in the field in numerous ways.

%% file: _sections/Conclusion/conclusion.tex
\section{Conclusion}

The previous sections identified numerous quantum control methods that extend the robustness of the primitive \msgate\ entangling gate. Three classes of gate schemes were first explored which add robustness to spin decoherence: pulsed, continuous and multi-level continuous dynamical decoupling. For each scheme, we characterized the trade-offs between the fidelity and the gate duration, the robustness to static frequency shifts, the added experimental complexity, and the calibration requirements. It was found that pulsed and continuous dynamical decoupling are very similar, and differ only slightly in their gate duration and fidelity. Multi-level continuous dynamical decoupling, however, results in a supperior gate fidelity and duration at the cost of a higher experimental overhead, additional fields, and more complex calibrations. 

Robustness to motional decoherence was found to be entirely attributed to the motion's path in phase space. Therefore, quantum control methods which extend the robustness to motional decoherence are schemes that enable an arbitrary phase space trajectory. The quality of a particular technique is then determined from its trade-off with the prolonged gate duration. Phase and frequency modulation of the sideband fields are found to be the most efficient quantum control schemes, as they result in the fastest gate for a given power budget. 

The quantum control schemes presented in this manuscript make up a library of tools that are available to the experimentalist. The various control methods can be combined with one another to obtain simultaneous robustness to multiple sources of decoherence. This is experimentally demonstrated by combining continuous dynamical decoupling with sideband phase modulation, achieving robustness to both spin and motional decoherence. 

%% file: _sections/Acknowledgments/acknowledgments.tex
\section{Acknowledgments}

 This work was supported by the UK Engineering and Physical Sciences Research Council via the EPSRC Hub in Quantum Computing and Simulation (EP/T001062/1), the UK Quantum Technology Hub for Networked Quantum Information Technologies (No. EP/M013243/1), the European Commission’s Horizon-2020 Flagship on Quantum Technologies Project No. 820314 (MicroQC), the US Army Research Office under Contract No. W911NF-14-2-0106, the Office of Naval Research under Agreement No. N62909-19-1-2116, the University of Sussex, and through a studentship in the Quantum Systems Engineering Skills and Training Hub at Imperial College London funded by the EPSRC (EP/P510257/1).

%% file: _sections/AppendixB/appendixB.tex
\section{Infidelity from off-resonant coupling to spectator motional states} \label{app:off_res_coupling_motion}

Off-resonant excitation of spectator states is investigated by calculating the expected infidelity from \eqref{eq:off_res_coupling_spectator_mode}. The residual displacement of the target mode is set to zero. An upper bound is placed on the infidelity by considering the maximal residual displacement of the spectator modes and setting $|\alpha_{j,k}(\tau)| \rightarrow \textrm{max}(|\alpha_{j,p}(t))$. Assuming that only the target mode is cooled to its ground state, the motional temperature of the spectator modes is taken from the Doppler cooling limit, $\nbar_p = \Gamma / 2\nu_p$ with $\Gamma/2\pi = \SI{19.6}{MHz}$. We first consider a set of experimental parameters corresponding to a smaller gradient with $\partial_zB = \SI{25}{T/m}$, $\nu_z/2\pi = \SI{220}{kHz}$ and $\Omega/2\pi = \SI{30}{kHz}$. For a two-ion chain and a gate performed on the COM (STR) mode, the expected infidelity is $\infid \approx \num{2e-7}$ ($\num{2e-5}$). Note that errors from off-resonant coupling to the COM mode are greater since the coupling strength is larger. Alternatively, for a higher gradient with $\partial_zB = \SI{150}{T/m}$, the resulting COM (STR) infidelity is $\infid \approx \num{7e-6}$ ($\num{7e-4}$). The infidelities are expected to grow for larger gradient strengths. Nevertheless, the expected fidelities reported here are well below the fault-tolerant threshold and can be made arbitrarily small by choosing an appropriate parameter regime (i.e. by increasing the secular frequency or decreasing the magnetic field gradient strength).

%% file: _sections/AppendixC/appendixC.tex
\section{Modulation sequences corresponding to a multi-tone MS gate} \label{app:mod_seq_mtms}

In section \ref{sec:engineering_arbitrary_pst} of the main text, sideband modulation is compared to multi-tone \msgate\ gates with respect to the gate time cost. That is, to obtain a robust phase space trajectory (PST), the path inherently covers a smaller area and there arises a trade-off with the gate time. In order to compare the schemes with one another, we find a modulation sequence for amplitude, phase and frequency modulation that matches the PST of an $N$-tone gate. This appendix presents the numerical methods as well as the obtained sequences.

We define $\alpha^{(N)}(t)$ as the PST of an $N$-tone gate. We are therefore interested in finding a discrete sequence of $\Omega_m$, $\delta_m$ or $\phi_m$ such that $\alpha(t) = \alpha^{(N)}(t)$. In what follows, an example for finding a phase sequence $\phi_m$ is presented, however this method is applicable to all types of modulation. We first define a time chunk $t_{chunk} = t_{m+1} - t_m$ much smaller than the gate time, $t_{chunk} \ll \tau_0$. The initial position of the PST is $\alpha(t=0) = 0$. The initial phase $\phi_0$ is found by minimizing $|\alpha(t_{chunk}, \phi_0) - \alpha^{(N)}(t)|$. 
This also allows us to find the time $t_0$ at which both PSTs intersect. The process is repeated and every $m^{th}$ step minimizes $|\alpha(m t_{chunk}, \phi_m) - \alpha^{(N)}(t)|$, with the condition that $t > t_m$. A dictionary of phases $\phi_m$ is built until the condition $t > \tau_0$ is met. The gate time of the modulation sequence is $m t_{chunk}$, and the gate time scaling is therefore $m t_{chunk} / \tau_0$.

\begin{figure}[t]
\center
\includegraphics[scale=1]{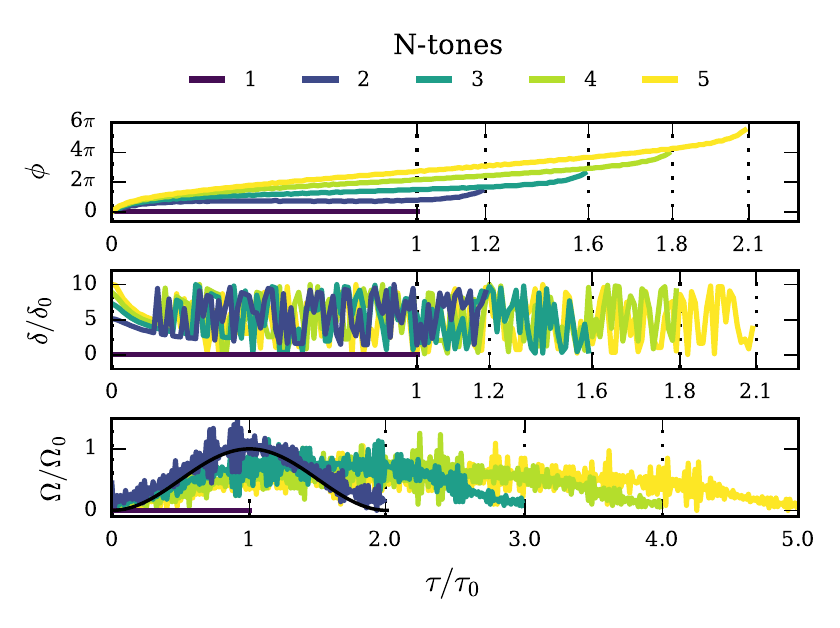} 
\caption{Modulation sequences for PM (top), FM (middle) and AM (bottom). The sequences are found numerically such that the resulting PST corresponds to an $N$-tone gate. The black line in the bottom plot corresponds to $\sin(\frac{\pi}{2} \frac{t}{\tau})^2$.}
\label{fig:mod_sequence}
\end{figure}

The resulting sequences are plotted in figure \ref{fig:mod_sequence}. The noise in the FM and AM parameters is due to the minimization step being done over a discrete sample. The sequences confirm that FM and PM sequences are more efficient as they result in faster gates for the same PSTs. This can be seen by examining the AM sequence, where the Rabi frequency is only a fraction of the total amplitude during the evolution. Since the AM sequences achieve a PST by effectively changing the gate speed, it is expected to result in less efficient gate times. An interesting scenario arises from the AM sequence corresponding to a 2-tone gate. The time-dependent Rabi frequency is well approximated by $\Omega(t) = \Omega_0 = \sin(\frac{\pi}{2} \frac{t}{\tau})^2$. This interestingly corresponds to the analytical amplitude pulse shaping derived in \cite{zarantonello2019}.

%% file: _sections/AppendixD/appendixD.tex
\section{Experimental details} \label{app:experimental_details}

The qubit is encoded in one of two $^{171}Yb^+$ ions that are confined within a macroscopic segmented Paul trap. The magnetic field at the center of the trap is $B_0 = \SI{7.4e-4}{T}$, and the strength of the magnetic field gradient is $\SI{23.6(3)}{T/m}$. Doppler cooling, state preparation and state readout are achieved via a 369 nm laser beam, however all coherent operations are performed using solely microwave and RF radiation. Probabilities are inferred from photon counts collected on a photo-multiplier tube and typical SPAM fidelities are 97\%, which are corrected for with a maximum log-likelihood method. 

The bichromatic fields are tuned closely to the stretch mode of motion whose frequency is $\nu/2\pi = \SI{382.6}{kHz}$. After doppler cooling, the stretch mode is further cooled by means of sideband cooling, which brings the average Fock state to $\nbar = \num{0.18(4)}$. The heating rate was measured to be $\ndot < \SI{0.7}{s^{-1}}$. The motional coherence times under a Ramsey and spin echo experiment are $T_2^* = \SI{57(5)}{ms}$ and $T_2 = \SI{0.99(22)}{s}$.

%% file: _sections/AppendixE/appendixE.tex
\section{Filter functions} \label{app:filter_functions}

As described in the main text, decoherence can be modelled in frequency space by a noise source's PSD that is subject to a transfer function. In this way, the qubit is made to act like a filter, whose filtering properties are directly computed from the pulse sequence. We here present several filter functions corresponding to the three classes of quantum control methods that extend the robustness to spin decoherence: PDD, CDD and MLCDD. 

\begin{figure}[t!]
\center
\includegraphics[scale=.65]{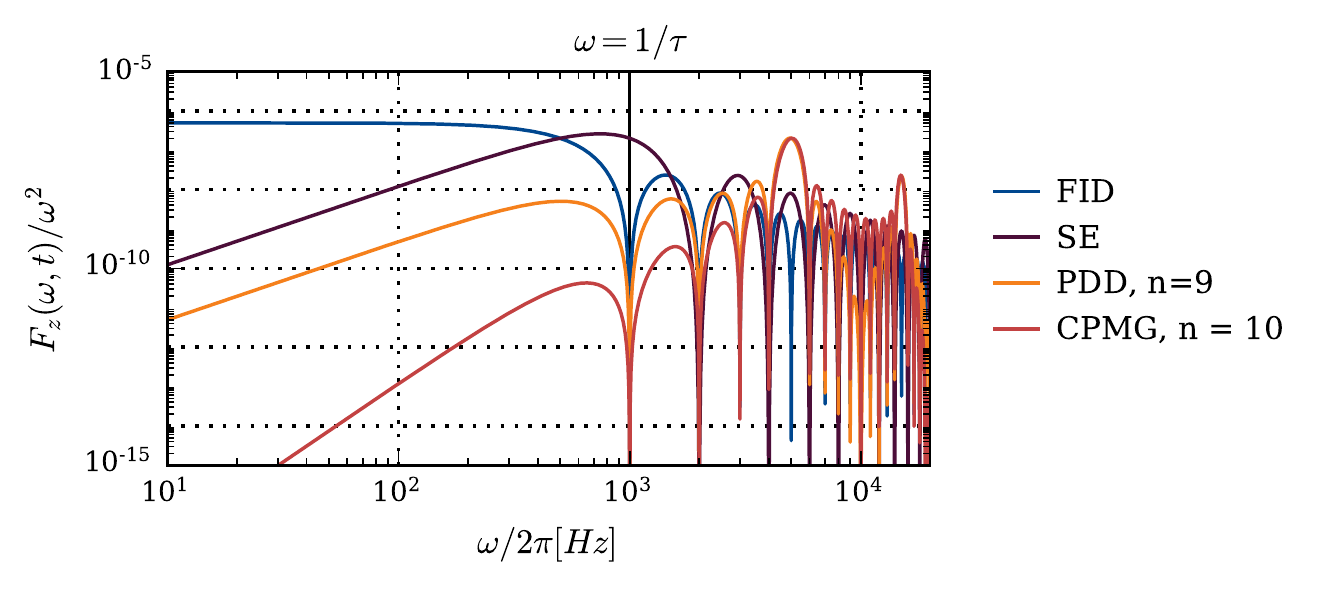} 
\caption{Filter functions corresponding to various dynamical decoupling sequences: Free Induction Decay (FID), Spin Echo (SE), Periodic Dynamical Decoupling (PDD) and Carr-Purcell-Meiboom-Gill (CPMG). The total duration is set to $\tau = \SI{1}{ms}$. The PDD and CPMG sequences are implemented with 9 and 10 $\pi$-pulses respectively.}
\label{fig:pdd_filter_function}
\end{figure}

The filter functions of various PDD sequences are plotted in figure \ref{fig:pdd_filter_function}. Under no dynamical decoupling, the qubit acts like a lowpass filter and is generally affected by most of the noise spectrum. The addition of $\pi$-pulses lowers the sensitivity of the interaction to low frequency noise. Furthermore, pulse timings such as those of CPMG \cite{carr1954, meiboom1958} result in a steeper roll-off than for periodic PDD, leading to an increased robustness.

\begin{figure}[t!]
\center
\includegraphics[scale=.65]{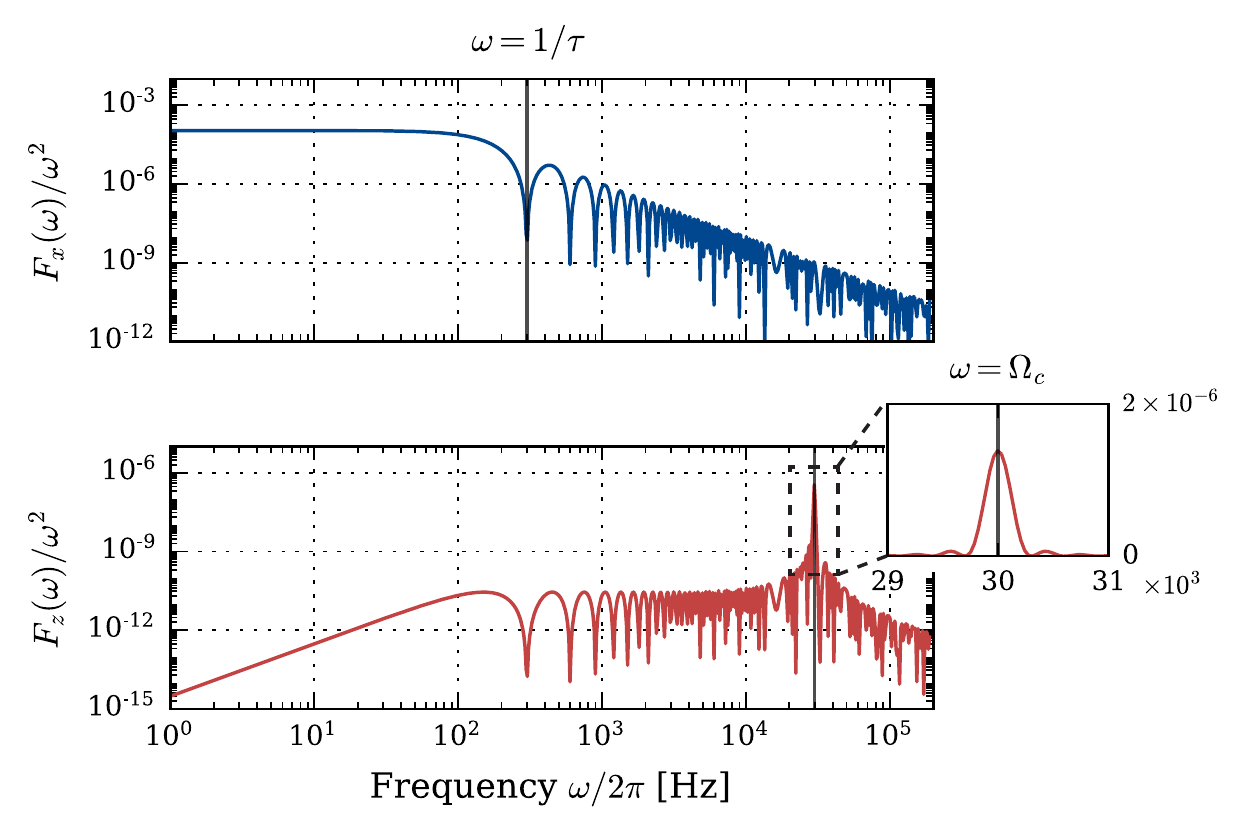} 
\caption{Filter functions from a driven evolution, i.e. continuous dynamical decoupling, where the Rabi frequency is set to $\Omega_c/2\pi = \SI{30}{kHz}$ and the total duration is $\tau = \SI{3.33}{ms}$. (Top) Amplitude filter function, which coincides with that of an FID, i.e. a low pass filter with a characteristic cutoff frequency $\omega/2\pi = 1/\tau = \SI{300}{Hz}$. (Bottom) Filter function for qubit frequency fluctuations. The zoomed in inset (linear scale) shows that it is well approximated by a sinc function centred at the carrier’s Rabi frequency.}
\label{fig:cdd_filter_function}
\end{figure}

\begin{figure}[t!]
\center
\includegraphics[scale=.65]{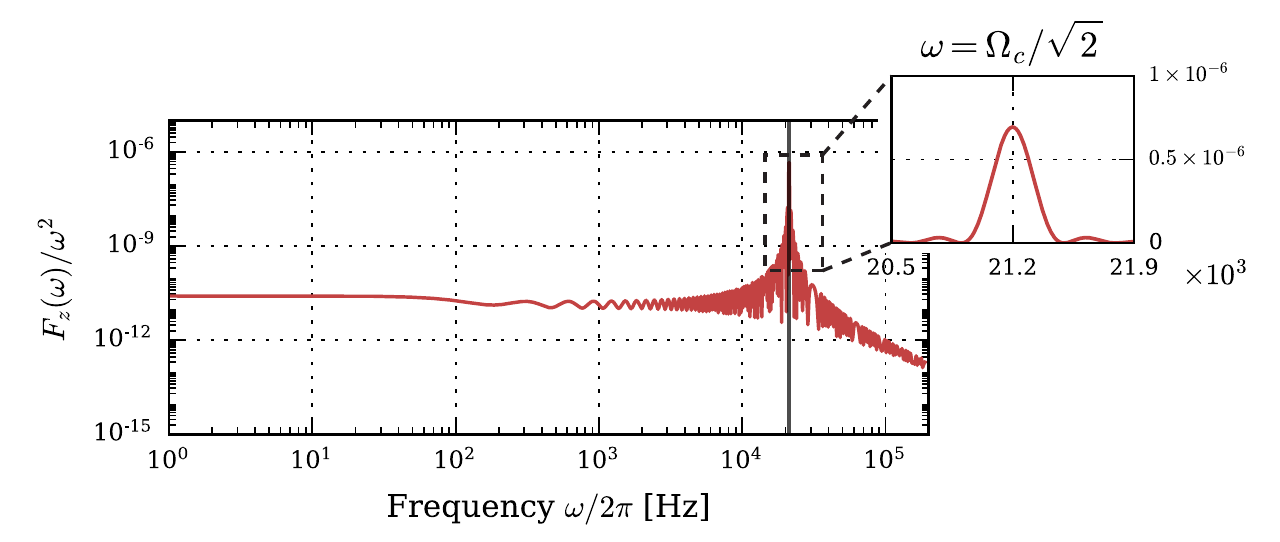} 
\caption{Filter function corresponding to qubit frequency noise for the dressed states, i.e. multi-level continuous dynamical decoupling. The system is modelled with the Hamiltonian of equation \ref{eq:mlcdd_hamiltonian_mw} of the main text, where $\Omega_c/2\pi = \SI{30}{kHz}$, and noise is included via equation \ref{eq:H_noise_mlcdd}. The inset shows that the filter function is well approximated by a sinc function that is centred around $\Omega_c/\sqrt{2}$.}
\label{fig:mlcdd_filter_function}
\end{figure}

The filter functions under CDD and MLCDD, computed from numerical simulations \cite{ball2021, qctrl_filter_function}, are presented in figures \ref{fig:cdd_filter_function} and \ref{fig:mlcdd_filter_function}. Both filters are well approximated by sinc functions centred around their respective bandpass frequency. Furthermore, their bandwidths are sufficiently small to justify the approximations of the filter functions by a Dirac-delta function in the main text. We also verify from figure \ref{fig:cdd_filter_function} that the primitive CDD scheme is not robust to amplitude noise in the drive, as the filter function is that of a free induction decay.

%% file: _sections/AppendixF/appendixF.tex
\section{Carrier off-resonant coupling during continuous dynamical decoupling} \label{app:cdd_off_res_coupling}

Continuous dynamical decoupling requires an additional continuous carrier field applied throughout the gate's evolution. Infidelities may arise from off-resonant coupling of this carrier field to the motional sidebands. In this appendix, we derive an approximate infidelity function for this error mechanism. Note that the derivations closely follow those presented in Ref. \cite{arrazola2020}.

Let us first consider the system's Hamiltonian describing the usual \msgate\ bichromatic field,

\begin{align}
& H = H_0 + H_{MS} + H_{c}, \nonumber\\
& H_0 = \sum_j \frac{\hbar\omega_0^{(j)}}{2} \sigma^{(j)}_z  + \hbar \nu a\dag a,  \nonumber \\
& H_{MS} = \hbar \epsilon \nu (a + a\dag)S_z,
\end{align}

with $S_i = \sigma_i^{(1)} + \sigma_i^{(2)}$. Considering an additional carrier field and moving into an interaction picture with respect to $H_0$, 

\begin{align} \label{eq:cdd_full_hamiltonian}
\tilde{H} = & \hbar\epsilon \nu (ae^{- i \nu t} + a\dag e^{i \nu t})S_z) \nonumber \\ 
& + \hbar\Omega_0 \cos(\delta t) S_x  + \frac{\hbar\Omega_c}{2}S_y ,
\end{align}

where $\Omega_0$ and $\delta$ are the Rabi frequency and detuning of the bichromatic fields, and $\Omega_c$ is the Rabi frequency of the carrier dynamical decoupling drive. In the bichromatic interaction picture rotating with $\hbar\Omega_0 \cos(\delta t)(\sigma^{(1)}_x + \sigma^{(2)}_x)$, equation \ref{eq:cdd_full_hamiltonian} becomes

\begin{align}
\tilde{H} = & \hbar \epsilon \nu (a e^{-i \nu t} + a\dag e^{i \nu t}) (J_0(\frac{2\Omega_0}{\delta})S_z + 2 J_1(\frac{2\Omega_0}{\delta})S_y ) \nonumber \\
& + \frac{\hbar\Omega_c}{2} \left( J_0(\frac{2\Omega_0}{\delta})S_y - 2J_1(\frac{2\Omega_0}{\delta})\sin(\delta t)S_z \right),
\end{align}

where $J_n(x)$ are Bessel functions of the first kind. The last term of equation describes off-resonant coupling of the carrier field to the sidebands with a detuning $\delta$ and Rabi frequency $\Omega_c\Omega_0/\delta$, where we've used $J_1(x) \approx x/2$. Using the supplementary material of Ref. \cite{khromova2012}, we approximate the infidelity by the transition probability from off-resonant coupling to be

\begin{equation} 
\infid = (1 + \frac{\delta^4}{\Omega_c^2\Omega_0^2})^{-1},
\end{equation}

which corresponds to equation \ref{eq:infidelity_cdd_off_res} of the main text under the assumption that $\delta \approx \nu$.